\documentclass[12pt]{article}
\usepackage[utf8]{inputenc}
\pdfoutput=1
\usepackage{jheppub}
\usepackage{setspace}
\usepackage{slashed}
\usepackage{graphicx,url}
\usepackage{booktabs}
\usepackage{bbm}
\usepackage{braket}
\usepackage{amsmath,hyperref,amssymb,cancel,stmaryrd}
\usepackage[usenames,dvipsnames]{xcolor}
\usepackage{color} 
 
\usepackage{epsfig} 
\def\half{{1 \over 2}}

\def\tr{{\rm Tr}}

\def\Or[#1]{{\text{O}}\left({#1}\right)}
\def\dotl[#1,#2]{\left\langle #1, #2 \right\rangle}
\def\dotlb[#1,#2]{[ #1, #2 ]}
\def\dotp[#1,#2]{(#1) \cdot (#2)}
\def\aff[#1,#2]{\hat{#1}(#2)}
\def\n4sym{{\cal N}=4 SYM}
\def\>{\rangle}
\def\<{\langle}
\def\weight[#1,#2,#3]{\{(#1),#2,#3\}}
\def\ads[#1]{$\text{AdS}_{#1}$}

\newcommand{\ba}{\begin{eqnarray}}
\newcommand{\ea}{\end{eqnarray}}
\newcommand{\be}{\begin{eqnarray}}
\newcommand{\ee}{\end{eqnarray}}
\newcommand{\bq}{\begin{equation}}
\newcommand{\eq}{\end{equation}}
\newcommand{\benn}{\begin{equation*}}
\newcommand{\eenn}{\end{equation*}}
\newcommand{\bi}{\begin{itemize}}  
\newcommand{\ei}{\end{itemize}}
\renewcommand{\d}{\partial}

\newcommand{\CI}{{\cal I}}
\newcommand{\CL}{{\cal L}}

\newcommand{\CN}{{\cal N}}
\newcommand{\CO}{{\cal O}}
\newcommand{\CP}{{\cal P}}

\newcommand{\nn}{\nonumber}
\renewcommand{\Im}[0]{\operatorname{Im}}
\renewcommand{\Re}[0]{\operatorname{Re}}

\newcommand\oo\infty
\newcommand\s\sigma
\newcommand\de\delta
\newcommand\De\Delta

\newcommand\f\phi
\newcommand\g\gamma
\newcommand\x\times

\newcommand\lrpar{\raise .8ex\hbox{$^\leftrightarrow$} \hspace{-9pt}
\partial}
\newcommand{\llb}{\llbracket}
\newcommand{\rrb}{\rrbracket}

\newcommand{\gp}{\mathfrak{q}}

\makeatletter
\def\@fpheader{\vspace{-.1cm}}
\makeatother

\title{
Conformal Truncation of Chern-Simons Theory at Large $N_f$
}

\author[a]{Luca V.~Delacr\'etaz,}
\author[b]{ A.~Liam Fitzpatrick,}
\author[b]{ Emanuel Katz,}
\author[b]{Lorenzo G.~Vitale}

\affiliation[a]{Department of Physics, Stanford University, \\
 Stanford, CA 94305, USA}

\affiliation[b]{Department of Physics, Boston University, \\
Commonwealth Avenue, Boston, MA 02215, USA}

\abstract{We set up and analyze the lightcone Hamiltonian for an abelian Chern-Simons field coupled to $N_f$ fermions in the limit of large $N_f$ using conformal truncation, i.e. with a truncated space of states corresponding to primary operators with dimension below a maximum cutoff $\Delta_{\rm max}$. In both the Chern-Simons theory, and in the $O(N)$ model at infinite $N$, we compute the current spectral functions analytically as a function of $\Delta_{\rm max}$ and reproduce previous results in the limit that the truncation $\Delta_{\rm max}$ is taken to $\infty$.   Along the way, we determine how to preserve gauge invariance and how to choose an optimal discrete basis for the momenta of states in the truncation space.  
}

\arxivnumber{}

\begin{document}

\maketitle  

%\tableofcontents

%######################################################################%
%======================================================================%
%======================================================================%
%======================================================================%
%######################################################################%
\section{Introduction and Summary}

Compared to the perturbative regime, the strong coupling regime of Quantum Field Theories (QFTs) remains poorly understood.  The standard method of defining QFTs nonperturbatively is through their Lattice definition, which puts the Lagrangian and the path integral front and center.  An alternative to the Lattice description are Hamiltonian truncation methods, which involve truncating the Hamiltonian of the theory to a finite subspace where it can be exactly diagonalized, usually numerically.  
A special case is to consider the large class of QFTs that are points along the RG flow of a Conformal Field Theory (CFT) deformed by a relevant operator.  In this case, if one fixes the truncation basis in terms of the primary operators of the original CFT, then the Hamiltonian matrix elements are given by the CFT OPE coefficients and its spectrum of operators.  We will refer to such methods generally as Conformal Truncation methods; these are an appealing formulation of QFT that takes CFTs rather than Lagrangians as the starting point.   

One of the major obstacles to applying conformal truncation widely is that there are few CFTs where the CFT `data' -- OPE coefficients and scaling dimensions -- are known to high precision.  Free theories are an obvious case where this data is known.  However, free theories are a problematic starting point for studying gauge theories. The reason is that interacting gauge theories are not really deformations of their free theory limit by a local operator;  $A_\mu J^\mu$ is not locally gauge
invariant.  Nevertheless, conformal truncation has been successfully applied to nonabelian gauge theories in $d=2$ \cite{Bhanot:1993xp, Katz:2013qua, Katz:2014uoa}.  An important simplification in this case is that 2d gauge fields have no local degrees of freedom, and can be integrated out of the action directly.  From this point of view, a natural next step is to consider Chern-Simons (CS) gauge theories coupled to matter in $d=3$, where again the gauge field is nondynamical and can be integrated out.  Understanding how to apply conformal truncation to such theories will be our main goal in this paper.  

Rather than considering CS gauge theories in general, we will restrict to a limit where the number $N_f$ of fermion flavors is infinite.  The large $N_f$ limit significantly simplifies the theory, and will allow us to perform all calculations analytically as well as to compare correlators to their known solution from resumming Feynman diagrams.  In addition, there are important conceptual issues that remain to be resolved at finite $N_f$, but are avoided at infinite $N_f$.

We will use the lightcone (LC) conformal truncation framework of \cite{Katz:2016hxp}. In general, there are tradeoffs between adopting an equal-time (ET) quantization scheme \cite{Yurov:1989yu, Hogervorst:2014rta, Rychkov:2014eea, Elias-Miro:2017tup} vs a LC quantization scheme \cite{Brodsky:1997de}.  In the present case, the most important reason we take LC quantization is that the interaction is dimensionless, and it remains unclear how to treat dimensionless interactions in ET Hamiltonian truncation.

Our motivation for starting a Hamiltonian Truncation study of CS gauge 
theories stems from several aspects. They are relevant to describe different 
physical phenomena, such as deconfined quantum criticality \cite{Senthil1490} and the $\nu=\half$ quantum Hall state \cite{PhysRevB.47.7312,Son:2015xqa}. Furthermore, CS theories with matter 
have been conjectured to undergo a web of dualities, which generalize 
rank-level duality \cite{Aharony:2015mjs,Karch:2016sxi,Seiberg:2016gmd}. Various evidence has been put forward 
to confirm these dualities, and we believe the Hamiltonian truncation could 
provide a strong non-perturbative check.

\subsection*{Summary of Results}

Our first task will be to construct the lightcone Hamiltonian in a basis of primary operators of the free, massless theory.  With the benefit of some hindsight, we will allow a mass counterterm for the gauge boson, which will eventually be tuned to cancel off the UV contributions due to a non-gauge-invariant regulator.  Aside from a term proportional to the gauge boson mass, the resulting Hamiltonian in terms of the fermions fields has appeared in the literature previously, e.g.~\cite{Giombi:2011kc}, and contains 2-, 4-, and 6-fermion terms,
\be
H= H_2 + H_4 + H_6.
\ee
  In lightcone quantization, these interactions can change fermion particle number by up to 0, 2, and 4 particles at a time, respectively.  However, in the large $N_f$ limit, fermion particle number is conserved and we can restrict to the two-particle sector.  The corresponding primary operators are therefore fermion bilinears, and our truncation scheme is to keep a basis of such operators up to a maximum scaling dimension, which in this case is equivalent to keeping operators up to a maximum spin $\ell_{\rm max}$.  
  
In this basis, diagonalizing the quadratic Hamiltonian $H_2$, which is just the Hamiltonian for the free massive theory, is already a nontrivial problem.  This is clearly true if the goal is to perform the diagonalization analytically, but even numerical diagonalization involves dealing with the fact that some of the matrix elements of $H_2$ in this basis are IR divergent.  To deal with this, we impose an IR regulator, and project onto the subspace that has finite energy as the IR regulator is taken to zero.  We refer to states in this space as the ``Dirichlet'' subspace.  Fortunately, it turns out to be possible to find this subspace and diagonalize $H_2$ acting on it analytically, for any value of the truncation level $\ell_{\rm max}$. We can then construct the spectral functions of local operators as sums over the overlaps with the resulting eigenstates.  For instance, we find the spectral function $\rho_{--}$ for the $\< J_- J_-\>$ current-current two-point function in the free theory is
\be
\pi \rho_{--}(q) &=&\frac{q_-^2}{4}\sum^{\frac{\ell_{\rm max}+1}{2}}_{\alpha=1} \frac{\frac{1}{\ell_{\rm max}+1} \cos^2 \frac{\pi (\alpha-\frac{1}{2})}{\ell_{\rm max}+1} }{\sqrt{q^2 - 4 m_f^2 \sec^2 \left( \frac{\pi(\alpha- \frac{1}{2})}{\ell_{\rm max}+1}\right)}},
\ee
where terms are only included in the sum if the argument of the square root in the denominator is positive.  Moreover, in the limit $\ell_{\rm max} \rightarrow \infty$, we recover the Lagrangian result,
\be
\pi \rho_{\mu\nu}(q)
	= \frac{q_\mu q_\nu - q^2\eta_{\mu\nu}}{|q|} \Re  \tau(q) - {\rm sgn}(m_f)\frac{i \epsilon_{\mu\nu\lambda}q^\lambda}{2\pi} \Im  \kappa(q)  , 
\ee
where $\tau$ and $\kappa$ are known functions.  

At large $N_f$, the only interaction terms that survive are in the four-body piece $H_4$ of the Hamiltonian.  Unlike the free Hamiltonian $H_2$, the interaction terms mix states with different invariant momentum-squared $\mu^2$, and the eigenstates become nontrivial functions of $\mu^2$. One way to deal with this mixing is simply to choose a priori a general basis, evaluate the matrix elements in this basis, and numerically diagonalize the resulting Hamiltonian.  The work in \cite{Katz:2016hxp} took such an approach, where the general basis was a particular fixed set of polynomials $g_k(\mu)$, vanishing sharply at $\mu$ above a cut-off $\Lambda$.  However, the more closely one can tailor the basis of eigenfunctions in $\mu$ to the interactions of the particular theory at hand, the smaller the basis that will be needed for a particular accuracy.  In the large $N_f$ limit, we will see that we can actually choose a small, optimal basis, selected by the structure of the interactions themselves.  The feature that makes this optimal choice possible is that the interaction matrix elements factorize into separate $\mu$-dependence of the bra and ket state.  As a consequence, we will be able to remove the cut-off analytically, and in fact we will see that we can take the truncation limit $\ell_{\rm max}$ to infinity as well, reproducing the known large $N_f$ correlators. 

 In fact, essentially the same strategy works to simplify the $O(N)$ model at large $N$.  We first treat this case as a warm-up, where we do not have to deal with issues of gauge invariance or spin indices. In the interacting theory, we again find we can analytically diagonalize the Hamiltonian, and so we obtain explicit expressions for the spectral functions at any $\ell_{\rm max}$.  In the $O(N)$ model, the spectral function for the $\phi^2$ two-point function takes the form
\begin{equation}
\pi
\rho_{\phi^2}(q) = \frac{\CP_+(q)}{(1+\frac{\lambda}{4} \CP_-(q))^2+ \frac{\lambda^2 }{16} \CP^2_+(q)}, \quad \CP_\pm(q) \equiv  \sum_{\alpha=1}^{\frac{\ell_{\rm max}}{2}} \frac{\frac{1}{\ell_{\rm max}+1}}{\sqrt{ \pm \left( q^2 - 4 m^2 \sec^2  \frac{\pi(\alpha-\frac{1}{2})}{\ell_{\rm max}+1}  \right) }}.
\end{equation}
 Again, the sum on $\alpha$ implicitly includes only terms where the argument of the square root is positive.  At infinite $\ell_{\rm max}$, this spectral function reduces to the Lagrangian result from summing bubble diagrams. Similarly, we obtain explicit analytic expressions for the spectral functions of the currents in the large $N_f$ Chern-Simons  theory, and find they reduce to the Lagrangian results at infinite $\ell_{\rm max}$.

The rest of the paper proceeds as follows.  In section \ref{sec:setup}, we set up the LC Hamiltonian for the abelian Chern-Simons gauge field coupled to fermions in lightcone gauge $a_-=0$,  and integrate out the nondynamical components of the fields.  In section \ref{sec:FreeTheory}, we construct the CFT primary basis states, and show how to analyze the free theory, where the only deformation of the UV CFT is the fermion mass term.  The analysis involves finding the eigenstates of the mass term in the truncated basis, and using these eigenstates to compute the spectral functions of the conserved currents.  In section \ref{sec:ON}, we warm up for the interacting CS case with an analysis of the $\CO(N)$ model at infinite $N$.  In section \ref{sec_CS_Nf}, we apply these methods to the Chern-Simons theory at infinite $N_f$, where we compute the spectral functions of the currents at finite $\ell_{\rm max}$ analytically.  We show that in the limit $\ell_{\rm max}\rightarrow \infty$ where the truncation level is taken to infinity, the correct results are reproduced.
Finally, in section \ref{sec:future}, we conclude with a discussion of potential future directions.

%######################################################################%
%======================================================================%
%======================================================================%
%======================================================================%
%######################################################################%
\section{Setup}
\label{sec:setup}

In this section, we set up the Lightcone (LC) Hamiltonian formulation of the theory of $N_f$ Dirac fermions coupled to a Chern-Simons gauge field.  Much of the construction will follow previous work (see e.g.~\cite{Giombi:2011kc}), though with some modifications in order to preserve gauge invariance once we pass to a truncation framework.  After removing the non-dynamical fields in light-cone gauge, the theory as formulated in \eqref{eq_L_final_ma} can be thought of as a deformation of a free fermion theory parameterized by a dimensionless coupling ($k$) and two relevant couplings ($m_f$ and $m_a$). The theory for general values of these couplings will be studied by diagonalizing the Hamiltonian in a truncated basis of free fermion states.

\subsection{Construction of On-Shell Hamiltonian}

We study $N_f$ Dirac fermions coupled to a $U(1)_{k}$ Chern-Simons gauge field
\begin{equation}\label{eq_theory}
\mathcal L 
	= \bar\Psi (i \cancel D_a + m_f)\Psi + \frac{k}{4\pi}\epsilon^{\mu\nu\lambda} a_\mu\d_\nu a_\lambda\, , 
\end{equation}
with $\cancel D_a \equiv \gamma^\mu(\d_\mu + ia_\mu)$.\footnote{Our choice of regulator will be such that integrating out the fermion in the $m_f\to \infty$ limit produces a shift $k\to k+\frac{N_f}{2}{\rm sgn}(m_f)$, and the theory is therefore gauge invariant if $k + N_f/2$ is an integer. However since we will mostly be concerned with the $N_f,k\to \infty$ limit, this distinction is unimportant here and we will treat $k$ as a free parameter in the following.}  This theory has a global% 
	\footnote{Technically, the global symmetry is a more complicated if one keeps track of discrete symmetries, see e.g.~\cite{PhysRevX.7.031051}.}	
$SU(N_f)$ flavor symmetry and a global $U(1)$ symmetry carried by the `topological' current
\begin{equation}\label{eq_U1_topo_current}
j^\mu = \frac{1}{2\pi} \epsilon^{\mu\nu\lambda} \d_\nu a_\lambda\, .
\end{equation}We will work in light-cone time ($x^+$) quantization with the following conventions 

\begin{equation}
x^{\pm}
	= \frac{1}{\sqrt{2}} (x^0\pm x^1)\, , \qquad
ds^2 
	= 2 dx^+ dx^- - dx_\perp^2\, , \qquad
\epsilon^{012} = -\epsilon^{+-\perp} = 1\,,
\end{equation}
and make the following choice of gamma matrices:
\begin{equation}
\gamma^+ = %\gamma_- = 
\left(\begin{matrix}0&\sqrt{2}\\0&0\end{matrix}\right)\, , \qquad
\gamma^- = %\gamma_+ = 
\left(\begin{matrix}0&0\\\sqrt{2}&0\end{matrix}\right)\, , \qquad
\gamma^\perp =
\left(\begin{matrix}-i&0\\0&i\end{matrix}\right)\, . \qquad
\end{equation}
We will work in light-cone gauge $a_-=0$. This gauge has a number of advantages, the chief one being that the gauge field is nondynamical and can be integrated out.  Naively, eliminating the gauge field also eliminates any issues with preserving gauge invariance, but unfortunately the actual situation is not so simple.  One way to think about the problem is that if we first integrate over the fermion degrees of freedom with a generic regulator, then we may generate a mass term for the gauge field. Defining the ``fermion current'' 
\be
j_f^\mu \equiv \bar \Psi \gamma^\mu \Psi,
\ee
 we can write
\be
\CL_{\rm eff}[a] &=&  \frac{k}{4\pi}\epsilon^{\mu\nu\lambda} a_\mu\d_\nu a_\lambda + a_\mu a_\nu \< j_f^\mu j_f^\nu\> 
 + \dots.
\ee
The two-point function $\<j_f^\mu j_f^\nu\>$ is linearly divergent and may produce a contribution $\sim \Lambda \eta^{\mu\nu}$ that generates a mass term for $a_\mu$.\footnote{Higher order terms are UV finite by Lorentz invariance and power-counting.} Consequently, our original Lagrangian may need a gauge boson mass counter-term to cancel this contribution:
\begin{equation}\label{eq:LagWithMa}
\mathcal L
	= \bar\Psi (i \cancel D_a + m_f)\Psi + \frac{k}{4\pi}\epsilon^{\mu\nu\lambda} a_\mu\d_\nu a_\lambda\, + m_a a_\mu a^\mu . 
\end{equation}
  In particular, when we attempt to diagonalize the lightcone Hamiltonian, we should expect to encounter divergences that are cancelled by a mass term for the gauge boson.  

Next, we want to integrate out the nondynamical fields from the Lagrangian.  First,  $a_+$ acts as a Lagrangian multiplier enforcing
\begin{equation}
a_\perp 
	= - \frac{2\pi}{k} \frac{1}{\d_-} j_f^+\, ,
\end{equation}
which allows us to eliminate $a_+$ and $a_\perp$.  The Lagrangian reduces to
\begin{equation}
\label{eq:LagNoA}
\mathcal L
	= \bar\Psi (i\cancel \d + m_f) \Psi + \frac{2\pi}{k}j_f^\perp \frac{1}{\d_-} j_f^+ + m_a  \left( \frac{2 \pi}{k} \right)^2 j_f^+ \frac{1}{\partial_-^2} j_f^+\, .
\end{equation}

The action can be further simplified because a component of the spinor is non-dynamical in light-cone quantization. 
Writing  the fermions as
$\Psi ={\chi\choose\psi}$, 
$\bar \Psi ={\psi^*\choose\chi^*}$, 
the currents are
\begin{equation}
j_f^+ = \sqrt{2} \psi^* \psi\, , \qquad
j_f^\perp = i (\chi^* \psi - \psi^*\chi)\, , \qquad
j_f^- = \sqrt{2} \chi^* \chi\, ,
\end{equation}
and the action for the components is
\begin{equation}
\begin{split}
\mathcal L
	&= i\sqrt{2} \left(\chi^* \d_-\chi + \psi^* \d_+ \psi \right) - \chi^*\d_\perp \psi + \psi^* \d_\perp \chi  \\
	& \qquad + m_f(\chi^* \psi+\psi^*\chi) + \frac{2\pi}{k} i\sqrt{2} (\chi^* \psi -\psi^*\chi) \frac{1}{\d_-} \psi^* \psi \\
	& \qquad + 2m_a \left(\frac{2\pi}{k}\right)^2 \psi^*\psi \frac{1}{\d_-^2} \psi^*\psi\, .
\end{split}
\end{equation}
The spinor component $\chi$ is non-dynamical, the solution to its constraint equation is
\begin{equation}\label{eq_chi_os}
\chi_{{\rm os} ,i}
	= \frac{1}{i\sqrt{2}} \frac{1}{\d_-} \left[\d_\perp \psi -m_f\psi_i - \frac{2\pi}{k}i\sqrt{2} \psi_i \frac{1}{\d_-} \psi_j^*\psi_j\right]\, .
\end{equation}
where we reintroduced the $SU(N_f)$ indices $i,j=1,\ldots,N_f$. After integrating out $\chi$, 
one obtains the following Lagrangian for $\psi$:
\begin{equation}\label{eq_L_final_ma}
\mathcal L
	= i\sqrt{2} \left(\psi^*_i \d_+ \psi_i - \chi_{{\rm os},i}^*\d_- \chi_{{\rm os},i}\right)
	+ 2m_a \left(\frac{2\pi}{k}\right)^2 \psi^*_i\psi_i \frac{1}{\d_-^2} \psi^*_j\psi_j\, ,
\end{equation}
This formulation of the theory will be our starting point for conformal truncation.

It will be useful to keep track of what has become of the global $U(1)$ current \eqref{eq_U1_topo_current} in the process of integrating out the non-dynamical fields. This can be done by coupling the theory \eqref{eq:LagWithMa} to a background gauge field:
\begin{equation}
\mathcal L [A] = \mathcal L + \frac{1}{2\pi}\epsilon^{\mu\nu\lambda}A_\mu \d_\nu a_\lambda\, .
\end{equation}
The current in the theory \eqref{eq_L_final_ma} without non-dynamical fields is then given by $j^\mu\equiv \delta S[A]/\delta A_\mu|_{A=0}$, or
\begin{equation}
\label{eq:JWithMa}
j^+ = \frac{1}{k} j_f^+\,, \qquad\qquad
j^\perp = \frac{1}{k} j_f^\perp + 2 m_a \frac{2 \pi}{k^2} \frac{1}{\d_-} j^+_f\,, 
\end{equation}
and $j^- = -\frac{1}{\d_-}(\d_+ j^+ + \d_\perp j^\perp)$. Note that the current is still identically conserved: $\d_\mu j^\mu=0$ holds whether or not the fermion $\psi$ is on-shell. This current conservation is in contrast with gauge invariance of the underlying theory, which holds only for a particular value of $m_a$. Moreover, because we are using a Lorentz violating gauge $a_-=0$, breaking of gauge invariance is tied to that of Lorentz invariance. Therefore, we can equivalently tune $m_a$ by requiring correlators of the purely fermionic theory \eqref{eq_L_final_ma} to be Lorentz invariant; this is what will be done in practice in section \ref{sec_CS_Nf}.

Following some straightforward manipulations outlined in appendix \ref{app_alg}, one can obtain the Hamiltonian $H$ in terms of the physical fermion modes from the above Lagrangian.  The Hamiltonian has two-particle, four-particle, and six-particle terms:
\be
H = H_2 + H_4 + H_6.
\ee
In LC quantization, the matrix elements of $H_2$ conserve total particle number because particles cannot be pair produced from the vacuum.  By contrast, $H_4$ ($H_6$) mixes $n$ and $n+2$ ($n, n+2$ and $n+4$) particle states.

%======================================================================%
%======================================================================%
%======================================================================%
\subsection{Simplifications at Infinite $N_f$}
\label{sec:infinitenf}

Up to this point, the construction has been for a general $N_f$.  We will now restrict to the case $N_f = \infty$, which has a number of simplifications.  The first of these simplifications is that particle number is conserved by the interaction at infinite $N_f$, and we can restrict our analysis to the two-particle states as these will not mix with higher particle numbers.   Moreover, we will focus on the singlet sector of $SU(N_f)$. Two-particle states in this sector take the form
\begin{equation}
\label{eq:singstates}
|\phi,P\rangle
	= \frac{A_\phi}{\sqrt{N_f}} \int d^3 y \, e^{-iyP} \left[f_\phi(-i\d^1,-i\d^2) :\psi^*(y_1) \psi(y_2): \right]_{y_{1,2}\to y} | 0 \rangle\, ,
\end{equation}
where here and in the following the summation over flavor indices will be kept implicit, and $A_\phi$ is a normalization factor.   The neutral states of the theory form a Generalized Free Field Theory (GFF), and they can be constructed from the two-particle eigenstates in a standard way. 

 At finite $N_f$, including particle-number-mixing interactions is a tractable challenge, but it requires numerically constructing as large a basis as possible.  By contrast, we will see that the infinite $N_f$ basis is simple enough to allow for a purely analytic treatment.

There are also conceptual issues that arise at finite $N_f$, having to do with the proper formulation of the Hamiltonian itself.  The first issue is that the double pole in \eqref{eq_L_final_ma}, if interpreted literally, leads to IR divergences in matrix elements.  A similar issue arises in the t' Hooft model.  In that case, one can derive the correct prescription for removing the IR divergence by analyzing Feynman diagrams in the covariant formulation, where the IR divergence cancels between multiple diagrams \cite{tHooft:1974pnl, Callan:1975ps}. A similar analysis might be possible here.  However, the issue is avoided entirely at infinite $N_f$, because the $t$-channel exchange of a photon vanishes in this limit.

A second issue is that the one-loop fermion propagator from integrating out the gauge field has a non-analytic contribution.  The one-loop contribution (and moreover the full resummed propagator at infinite $N_c$) was computed in \cite{Giombi:2011kc} in lightcone gauge and found to be
\be
\label{eq:FermionSigmaOneLoop}
\Sigma(p) = \frac{N_c}{k} \sqrt{2p_+ p_-}.
\ee
In terms of the loop integral of the rainbow diagram in Lorentzian space, this nonanalytic contribution comes entirely from the zero lightcone momentum $q_-=0$ region of the photon.  When we integrate out the photon in our LC Lagrangian, we generate a four-fermion term, and normal-ordering this four-fermion term involves the same contractions in the rainbow diagram.  However, it is not clear how the nonanalytic contribution from the rainbow diagram ought to be evaluated in the LC Hamiltonian.  A naive interpretation would be simply to evaluate  (\ref{eq:FermionSigmaOneLoop}) on-shell, effectively setting $p_+$ to the LC energy of the corresponding individual parton. On the other hand, generally nonanalytic terms must correspond to sums over physical states in the theory and cannot be absorbed into local terms in the action.  The infinite $N_c$ limit provides a promising setting in which to analyze this issue further, since the theory is solvable there \cite{GurAri:2012is} in the covariant approach.  However, rainbow diagrams vanish in the infinite $N_f$ limit.

Given these additional complications, we will address the problem at finite $N_f$ in future work and focus here on the limit $N_f,k,m_a\to \infty$, keeping the following ratios fixed
\begin{equation}
\lambda = \frac{N_f}{k}\, , \qquad\qquad
m_a' = \frac{m_a}{N_f}\, .
\end{equation}
We drop the prime on $m_a$ in the following for simplicity.

%######################################################################%
%======================================================================%
%======================================================================%
%======================================================================%
%######################################################################%
\section{Free Theory Warm-up} 
\label{sec:FreeTheory}

In the conformal truncation framework of Ref.~\cite{Katz:2016hxp}, only states that map to primaries of the free fermion CFT need to be considered.  To set up this basis, it will therefore be sufficient to consider only the free part of the mass matrix
\begin{equation}
\mathcal M \equiv P^\mu P_\mu = 2P_- H\, ,
\end{equation}
where the light-cone Hamiltonian is simply (see appendix \ref{app_H_modes})
\begin{equation}\label{eq_H_free}
H = H_2= \int \frac{d^2 p}{4\pi^2} \, \frac{p_\perp^2 + m_f^2}{2p_-} \, \psi^*(p) \psi(-p)\, .
\end{equation}
 A number of interesting issues arise already in the free theory.  For the sake of clarity, in this section we will therefore set up the conformal truncation basis and compute the current correlators in the free theory.  We will separate the quadratic Hamiltonian into a massless ``kinetic'' piece and a mass term, $H_2 = H^{\rm kin} + H^{\rm mass}$:
 \be
 H^{\rm kin} = \int \frac{d^2 p}{4\pi^2} \, \frac{p_\perp^2 }{2p_-} \, \psi^*(p) \psi(-p)\, , \qquad 
H^{\rm mass} = \int \frac{d^2 p}{4\pi^2} \, \frac{ m_f^2}{2p_-} \, \psi^*(p) \psi(-p)\, .
\label{eq:Hfree}
\ee
We will start by the describing the conformal basis that diagonalizes the kinetic term $H^{\rm kin}$, and then we will consider the mass term.

%======================================================================%
%======================================================================%
%======================================================================%
\subsection{Massless primary basis}

The 2-body primary operators of the free fermion CFT are spanned by the singlet $J^0 = \bar\Psi\Psi$ and higher spin currents $J^\ell_{\mu_1 \cdots \mu_\ell}$ for $\ell=1,2,\ldots$, which are known in closed form \cite{Giombi:2017rhm}. Ward identities reduce the number of independent components to 2 per current; we take these to be $J^\ell_{-\ldots--}$ and $J^\ell_{-\ldots-\perp}$, which are parity even and odd respectively. These are shown in appendix \ref{app_higherspin} to lead to the states 
\begin{equation}
\label{eq:masslessbasis}
\begin{split}
|\ell,s,\mu\rangle
	&= \frac{A}{\sqrt{N_f}}
	\int_0^1 \frac{dx}{\pi} \, \sqrt{x(1-x)} \frac{1}{2}\sum_{\sigma=\pm} f_{\ell s}(x,\sigma) :\psi^*_p \psi_{P-p}: | 0 \rangle\, ,
\end{split}
\end{equation}
Here, $\ell$ denotes the spin of the state, $s=\pm$ denotes the sign under parity, and $\mu$ is the $P^2$ eigenvalue.  We have also introduced the normalization factor $A = \sqrt{\frac{P_-}{\mu}}$. The wavefunctions $f_{\ell s}$ are
\begin{equation}\label{eq_app_f_ell}
f_{\ell +}(x,\sigma) = 2C_{\ell-1}^{1}(1-2x)\, , \qquad \qquad
f_{\ell -}(x,\sigma) = \frac{\sigma \ell/2}{\sqrt{x(1-x)}}C_{\ell}(1-2x)\, ,
\end{equation}
for $\ell\geq 1$, and the singlet is given by
\begin{equation}\label{eq_app_f_0}
f_{0-}(x) 
	= \frac{\sigma/\sqrt{2}}{\sqrt{x(1-x)}}\, .
\end{equation}
These states are properly normalized (see appendix \ref{app_matrixel}):
\begin{equation}\label{eq_normalize_states}
\langle \ell',s',\mu' | \ell,s,\mu\rangle = 2\pi \delta(\mu^2 - \mu'^2) \delta_{\ell\ell'}\delta_{ss'}\, .
\end{equation}
The matrix elements of the kinetic term $H^{\rm kin}$ from (\ref{eq:Hfree}) are diagonal and given by
\begin{equation}
\label{eq:kinterm}
\mathcal M_{\ell's',\,\ell s}^{\rm kin}
	= 2\pi \delta(\mu^2 - \mu'^2)\delta_{\ell \ell'}\delta_{ss'} \cdot \mu^2\, .
\end{equation}
%

%======================================================================%
%======================================================================%
%======================================================================%
\subsection{Massive theory and IR Divergences as Projectors: The Dirichlet Basis}
When $m_f\neq 0$, the parity even sector has the following matrix elements for the mass term $H^{\rm mass}$ in \eqref{eq:Hfree}:
\begin{equation}\label{eq_masselements_peven}
\mathcal M^{\rm mass}_{\ell' s,\, \ell s}
	= 2\pi \delta(\mu^2 - \mu'^2)\delta_{\ell \ell' {\rm mod} 2}\delta_{ss'} \cdot M^s_{\ell \ell'}
	\qquad \hbox{with} \qquad  M^+_{\ell \ell'} = 8m_f^2 \min(\ell,\ell')\, .
\end{equation}
Notice that spin is conserved mod 2. One can easily check that $M^+$ has a spectrum that is bounded below by $4m_f^2$ as expected. The parity odd sector however has IR divergences: 
\begin{equation}\label{eq_M_IRdiv}
M^-_{\ell\ell'} = \frac{8m_f^2 }{2^{(\delta_{\ell 0}+\delta_{\ell' 0})/2}}\left[\frac{1}{\sqrt{\epsilon}} - \max(\ell,\ell')\right]\, ,
\end{equation}
where $\epsilon$ is a sharp IR momentum cutoff, see appendix \ref{app_dirichlet} for details. In appendix \ref{sec:DimRegDirichlet}, we discuss an IR regulator based instead on dimensional regularization.  The spectrum of $M^-$ is also bounded by $4m_f^2$, but has 4 eigenvalues that are strictly infinite in the limit $\epsilon\to 0$. The corresponding states will thus be projected out and will not contribute to spectral densities. It is then convenient to eliminate these states from the start, by projecting the basis obtained in the previous section to the kernel of the IR divergent part of $M^-$. This is done in appendix \ref{app_dirichlet} and leads to states of the form \eqref{eq:masslessbasis} with now
\begin{equation}
f_{\ell -}(x,\sigma)
	= \sigma  \frac{\sqrt{x(1-x)}}{\sqrt{\ell^2-1}}8C_{\ell-2}^2(1-2x)
	\qquad \hbox{for} \quad \ell\geq 2\, .
\end{equation}
The matrix elements still have the form \eqref{eq_masselements_peven}, but now with
\begin{equation}
\label{eq:MmassDirichlet}
M_{\ell\ell'}^-
	= \frac{8}{3} m_f^2 \min(\ell,\ell') \sqrt{\frac{\min(\ell,\ell')^2-1}{\max(\ell,\ell')^2-1}}\, .
\end{equation}
The parity even sector is unchanged since $M^+$ is free of IR divergences.

%======================================================================%
%======================================================================%
%======================================================================%
\subsection{Diagonalization of the mass term}
\label{sec:MassDiagonalization}

The mass matrix \eqref{eq_masselements_peven} is block diagonal in four sectors labelled by parity and spin mod 2, each of which can be diagonalized independently. We focus here on the spin odd sector, and treat the spin even sector in appendix \ref{app_spin_even}.  Our goal will be to find the eigenbasis of the massive quadratic Hamiltonian in the truncated conformal basis.  Because the kinetic term is diagonal, diagonalizing the massive quadratic Hamiltonian $H_2$ is equivalent to diagonalizing the mass term $H^{\rm mass}$.   We will use Greek letters $|\alpha\>, |\beta\>$ to refer to parity even and odd eigenstates, respectively, of the (truncated) quadratic Hamiltonian, whereas $|\ell,\pm\>$ denotes the parity even/odd conformal primary eigenstates of the kinetic term.

Let us start with the parity-even, spin odd sector, truncated to $\frac{1}{2}(\ell_{\rm max}+1)$ states
\begin{equation}
|1, +\rangle,\, 
|3, +\rangle,\, \ldots,\,
|\ell_{\rm max}, +\rangle.
\end{equation}
The Hamiltonian is diagonalized $\mathcal M | \alpha \rangle = \mathcal M_\alpha | \alpha \rangle$ by the following states:
\begin{equation}
\label{eq_evec_sodd_peven}
|\alpha\rangle = \sum_{j=1}^{\frac{\ell_{\rm max}+1}{2}} e^{(\alpha)}_{j+} | 2j-1,+ \rangle\, ,
\end{equation}
$\alpha=1,2,\ldots ,\frac{\ell_{\rm max}+1}{2}$, where
\begin{equation}
e_{j+}^{(\alpha)}
= \frac{2}{\sqrt{\ell_{\rm max}+1}} (-1)^{j} \cos \left[ 2 \phi_{\alpha+} \left(j - \frac{1}{2}\right)\right] \, , \quad \phi_{\alpha+} \equiv  \frac{\pi \left(\alpha-\frac{1}{2}\right)}{\ell_{\rm max}+1},
\end{equation}
and with eigenvalues
\begin{equation}\label{eq_eval_sodd_peven}
\mathcal M_{\alpha+} =
	%\equiv 
	\mu^2 + 4m_f^2 \sec^2 \phi_{\alpha+}\, .
\end{equation}
Next we consider the parity-odd, spin odd sector truncated to $\frac{1}{2}(\ell_{\rm max}-1)$ states
\begin{equation}
|3, -\rangle,\, 
|5, -\rangle,\, \ldots,\,
|\ell_{\rm max}, -\rangle.
\end{equation}
In this sector, the Hamiltonian is diagonalized  $\mathcal M | \beta \rangle = \mathcal M_{\beta-} | \beta \rangle$ by  states of the form
\begin{equation}
\label{eq_evec_sodd_podd}
|\beta\rangle = \sum_{j=1}^{\frac{\ell_{\rm max}-1}{2}} e^{(\beta)}_{j-} | 2j+1,- \rangle\, ,
\end{equation}
$\beta=1,2,\ldots ,\frac{\ell_{\rm max}-1}{2}$.  We will not need the explicit form of the vectors $e^{(\beta)}_{j-}$, but we will need the explicit form of the eigenvalues:
\begin{equation}\label{eq_eval_sodd_peven}
\mathcal M_{\beta-} 
	\equiv \mu^2 + 4m_f^2 \sec^2 \phi_{\beta-}, \qquad \phi_{\beta-} \equiv \frac{\pi \beta}{\ell_{\rm max}+1}. 
\end{equation}
Note that the spectrum satisfies
\begin{equation}
\mathcal M_{\gamma \pm} > \mu^2 + 4 m_f^2
\qquad \hbox{and}  \qquad  
\lim_{\ell_{\rm max}\to \infty} \min_{\gamma}\mathcal M_{\gamma\pm} = \mu^2 + 4m_f\, ,
\end{equation}
with $\gamma = \alpha,\beta$.

%======================================================================%
%======================================================================%
%======================================================================%
\subsection{Massive free fermion correlators}
\label{sec:FreeFermionCorrelators}

As an application of the diagonal basis constructed in the previous section we can compute spectral densities
\begin{equation}\label{eq_spec_dens1}
\rho_{\mathcal{O}_1\mathcal{O}_2}(q)
	= \sum_i \delta(\mathcal M_i - q^2) \langle \mathcal{O}_1|i, q\rangle \langle i,  q | \mathcal{O}_2\rangle\, .
\end{equation}
in the free fermion theory. Here, the sum over $i$ is an integral over $\mu^2$ and a sum over primaries. If the operators $\mathcal{O}_{1,2}$ only overlap with the spin odd, two-particle primaries, the spectral density takes the form
\begin{equation}
\rho_{\mathcal{O}_1\mathcal{O}_2}(q)
	= 2q_-\sum_{\gamma} \int \frac{d\mu^2}{2\pi} \delta(\mathcal M_\gamma - q^2) \langle \mathcal{O}_1|\gamma,\mu^2,\vec q\rangle \langle\gamma,\mu^2,\vec q | \mathcal{O}_2\rangle\, ,
\end{equation}
where the states $|\gamma,\mu^2,\vec q\rangle$ were defined in Eqs.~\eqref{eq_evec_sodd_peven} and \eqref{eq_evec_sodd_podd}, with $\mu^2$ and $\vec{q}$ implicit (e.g. ``$|\alpha\>$'' $\equiv |\alpha, \mu^2, \vec{q}\>$) and the sum over $\gamma$ includes both parity even and parity odd states.
Given the simple form of the eigenvalues \eqref{eq_eval_sodd_peven}, we can perform the integral over $\mu^2$ to obtain
\begin{equation}\label{eq_spec_dens2}
\rho_{\mathcal{O}_1\mathcal{O}_2}(q)
	= \frac{q_-}{\pi}\sum_{\gamma} \Theta(\mu_\pm^2 (\gamma,q)) \langle \mathcal{O}_1|\gamma,\mu_\pm^2(\gamma,q),\vec q\rangle \langle \gamma,\mu_\pm^2(\gamma,q),\vec q | \mathcal{O}_2\rangle\, ,
\end{equation}
where $\mu_\pm^2 (\gamma,q) =q^2 - 4m_f^2 \sec^2 \phi_{\gamma\pm}$. The sum is over all $\gamma$'s small enough such that $\mu_\pm^2(\gamma,q)\geq 0$.

We will take the operators to be the currents $\mathcal{O} = j_-,\, j_\perp$. Since current is conserved, correlators involving these components entirely fix correlators involving $j_+$. These currents differ from the currents used to define the basis in two ways. First, the physical currents have a natural normalization $j_\mu = \bar\Psi\gamma_\mu \Psi$ so that the charge is an integer, whereas the higher spin currents used in the construction of the basis were normalized according to \eqref{eq_normalize_states}. Second, the physical currents of the interacting (or massive) theory do not necessarily match the ones defined in the free CFT. This is the case in particular for $j_\perp = -i(\chi^*\psi - \psi^* \chi)$, since the solution to the constraint equation for $\chi$ \eqref{eq_chi_os} depends on $m_f,\, m_a,\, k$. In the free case with only $m_f\neq 0$, we have
\begin{equation}
j_- = \sqrt{2} \psi^* \psi\, , \qquad
j_\perp = \frac{1}{\sqrt{2}} \psi^* \frac{\d_\perp-m_f}{\d_-} \psi + \hbox{h.c.}\, .
\end{equation}
The first step in obtaining spectral densities is to compute the overlap of the states of interest $|\mathcal{O}\rangle \equiv \mathcal{O}(y=0)|0\rangle$ with the basis states. Using Wick contractions one finds that the overlaps with the Dirichlet states are
\begin{subequations}
\begin{align}
\langle j_-| \ell,s,q\rangle
	&= \frac{\sqrt{q_-}}{4\sqrt{\mu}} \delta_{\ell 1} \delta_{s+}\, ,\\
\langle j_\perp | \ell,s,q\rangle
	&= \frac{-im_f}{\sqrt{\mu q_-}} \delta_{\ell 1{\rm mod }2}\delta_{s+}  + \frac{\sqrt{\mu}}{2\sqrt{q_-}} \frac{\delta_{\ell 1{\rm mod }2}}{\sqrt{\ell^2-1}}\delta_{s-}\, .
\end{align}
\end{subequations}
The overlaps with the parity-even Hamiltonian eigenstates are therefore
\begin{subequations}\label{eq_j_alpha}
\begin{align}
\langle j_-| \alpha,q\rangle
	&= \frac{\sqrt{q_-}}{4\sqrt{\mu}}  e_1^{(\alpha)} 
	= -\frac{\sqrt{q_-}}{2\sqrt{\mu}} \frac{\cos \phi_{\alpha+}}{\sqrt{\ell_{\rm max}+1}} \, ,\\
\langle j_\perp| \alpha, q\rangle
	&= \frac{-im_f}{\sqrt{\mu q_-}} \!\sum_{j=1}^{\frac{\ell_{\rm max}+1}{2}}e_j^{(\alpha)}\!
	= \frac{im_f}{\sqrt{\mu q_-}} \frac{\sec \phi_{\alpha+}}{\sqrt{\ell_{\rm max}+1}} \, ,
\end{align}
\end{subequations}
where $\phi_{\alpha+} = \frac{\pi \left(\alpha - \frac{1}{2}\right)}{\ell_{\rm max}+1}$. The component $j_\perp$ overlaps with parity-odd states:
\begin{equation}\label{eq_j_beta}
\langle j_\perp(0) | \beta,q\rangle
	= \frac{\sqrt{\mu}}{4\sqrt{q_-}}  \sum_{j=1}^{\frac{\ell_{\rm max}-1}{2}} \frac{e_j^{(\beta)} }{\sqrt{j(j+1)}}
	= \frac{\sqrt{\mu}}{2\sqrt{q_-}}  
	 \frac{(-1)^\beta}{\sqrt{\ell_{\rm max}+1}} \sin \phi_{\beta-} \, ,
\end{equation}
where $\phi_{\beta-} = \frac{\pi \beta}{\ell_{\rm max}+1}$. Consequently, parity-odd states only appear in the cut for $\rho_{\perp\perp}$. Note that Lorentz invariance is broken by truncation in the Dirichlet basis. Indeed, Lorentz generators $L_{\mu\nu}$ can rotate $|J_-\rangle $ into $|J_\perp\rangle$ and so on -- in the primary basis truncation therefore keeps or removes entire Lorentz multiplets without mutilating any. However in the Dirichlet basis $|J_-\rangle = |1,+\rangle$ is rotated onto all $|\ell,-\rangle$, so $L_{+\perp}|J_-\rangle$ is not in the truncated Hilbert space and Lorentz invariance is broken. This breaking of Lorentz invariance at finite $\ell_{\rm max}$ can be probed by studying $\rho_{\perp\perp}$, as discussed below. 

Plugging the overlaps \eqref{eq_j_alpha} into \eqref{eq_spec_dens2} one finds
\begin{subequations}\label{eq_spec_dens_sim2}
\begin{align}
\rho_{--}(q)
	&= \frac{1}{\ell_{\rm max} + 1} \sum_\alpha \llbracket \mu_+^2(\alpha,q)\rrbracket^{-1/2} \cos^2 \phi_{\alpha+} \cdot \left(\frac{q_-^2}{4\pi} \right)\, ,\\
\rho_{\perp -}(q)
	&=  \frac{1}{\ell_{\rm max} + 1} \sum_\alpha \llbracket \mu_+^2(\alpha,q)\rrbracket^{-1/2} \cdot \left(-\frac{im_f q_-}{2\pi}\right)\, , \\
\rho_{\perp\perp}(q)
	&= \frac{1}{\ell_{\rm max}+1} \cdot \\
	&\left[\sum_\alpha \llbracket \mu_+^2(\alpha,q)\rrbracket^{-1/2} \sec^2 \phi_{\alpha+} \cdot\left(\frac{m_f^2}{\pi}\right)  +  \sum_\beta \llbracket \mu_-^2(\beta,q)\rrbracket^{1/2} \sin^2 \phi_{\beta-}\cdot \left(\frac{1}{4\pi}\right)\right]\, , \nonumber
\end{align} 
\end{subequations}
where we introduced the short-hand notation $\llbracket x\rrbracket^{\alpha} \equiv x^\alpha\Theta(x)$. The spectral densities therefore take the form 
\begin{equation}\label{eq_rho_jj}
\pi \rho_{\mu\nu}(q)
	= \frac{q_\mu q_\nu - q^2\eta_{\mu\nu}}{|q|} \Re \tilde \tau(\gp) - {\rm sgn}(m_f)\frac{i \epsilon_{\mu\nu\lambda}q^\lambda}{2\pi} \Im \tilde \kappa(\gp) + \pi\tilde \rho_{\mu\nu}(q)\, , 
\end{equation}
where $\tilde \rho_{\mu\nu}(q)$ is the Lorentz violating part (with our kinematics $q_\perp = 0$ we have $\tilde \rho_{\mu\nu} \propto \delta_\mu^\perp\delta_\nu^\perp$)
and where
\begin{subequations}
\begin{align}
\Re \tilde \tau(\gp) 
	&= \frac{\gp/4}{\ell_{\rm max}+1} \sum_\alpha \llbracket \gp^2-{4}{} \sec^2 \phi_{\alpha+}\rrbracket^{-1/2} \cos^2 \phi_{\alpha+}\, , \\
\Im \tilde \kappa(\gp)
	&= \frac{\pi}{\ell_{\rm max}+1} \sum_\alpha \llbracket \gp^2-{4}{} \sec^2 \phi_{\alpha+}\rrbracket^{-1/2}  \, .
\end{align}
\end{subequations}
In the $\ell_{\rm max}\to \infty$ limit, one recovers the Lagrangian result \cite{Closset:2012vp}
\begin{equation}\label{eq_jj}
\frac{1}{N_f}\langle j^\mu j^\nu (q)\rangle_0
	= \tau_0(\gp) \frac{q^\mu q^\nu - q^2 \eta^{\mu\nu}}{|q|} + {\rm sgn\,}(m_f)\, \kappa_0(\gp) \frac{\epsilon^{\mu\nu\lambda}q_\lambda}{2\pi}\, , 
\end{equation}
with $\gp = |q|/|m_f|$ and 
\begin{subequations}
\label{eq:taukappa0}
\begin{align}
\kappa_0
	&= -\frac{1}{\gp} \tanh^{-1}\frac{\gp}{2}\, , \\
\tau_0
	&= -\frac{i}{4\pi \gp} - \frac{i(4+\gp^2)}{8\pi \gp} \kappa_0\, .
\end{align}
\end{subequations}
Indeed, 
\begin{subequations}
\begin{align}
\lim_{\ell_{\rm max}\to \infty} \Re \tilde \tau (\gp)
	&= \frac{\gp}{4}\int_0^{\frac{1}{\pi}{\rm acos} \frac{2}{\gp}}\! dx  \frac{\cos^2 \pi x}{\sqrt{\gp^2 - 4 \sec^2 \pi x}}
	=\frac{4 + \gp^2}{16 \gp^2} 
	= \Re \tau_0 (\gp)\, , \\
\lim_{\ell_{\rm max}\to \infty} \Im \tilde \kappa (\gp)
	&= {\pi}\int_0^{\frac{1}{\pi}{\rm acos} \frac{2}{\gp}}\! dx  \frac{1}{\sqrt{\gp^2 - 4 \sec^2 \pi x}}
	=\frac{\pi}{2 \gp} 
	= \Im \kappa_0 (\gp) \, ,
\end{align}
\end{subequations}
and the Lorentz violating part vanishes
\begin{equation}
\lim_{\ell_{\rm max}\to \infty}\tilde \rho_{\perp\perp}(q)
	= \frac{m_f}{4\pi}\int_0^{\frac{1}{\pi}{\rm acos} \frac{2}{\gp}} dx\, \frac{4 - \gp^2 \cos 2\pi x}{\sqrt{\gp^2 - 4 \sec^2 \pi x}} = 0\, .
\end{equation}
One can similarly recover stress tensor spectral densities which instead involve spin even states -- this is done in appendix \ref{app_spin_even}.

In the above expressions, we have kept track of the $q_-$-dependence, but in the remaining sections we will use boost invariance to set $q_-=1$.  

%######################################################################%
%======================================================================%
%======================================================================%
%======================================================================%
%######################################################################%
\section{$\CO(N)$ Warm-up} 
\label{sec:ON}

 Having seen how to diagonalize the free theory, we next want to diagonalize the interacting Hamiltonian.  The free Hamiltonian was diagonal in $\mu^2$, and therefore the energy eigenstates were also eigenstates of $\mu^2$.  Once we turn on interactions, the Hamiltonian will mix states with different $\mu^2$s, so we will have to contend with the Hamiltonian in a basis both of primary operators and a basis of $\mu$-dependence of the wavefunctions.  Our first step will be to choose an optimal basis for the $\mu$ dependence of the wavefunctions.  This basis will in turn allow us to solve for the eigenstates analytically, both in terms of their $\mu$ dependence and their components for the CFT primary operators.

In this section, we will go through these manipulations in a model that is simpler than Chern-Simons at large $N_f$, but similar in many ways: namely, the  3d scalar $\CO(N)$ model at large $N$.  This warm-up will allow us to demonstrate how to choose a basis to efficiently deal with the interactions, without having to also deal with issues relating to gauge invariance.  In the following we set $q_-=1$.

%======================================================================%
%======================================================================%
%======================================================================%
\subsection{Hamiltonian and Matrix Elements}
\label{sec:onmodel}

The Lagrangian of the theory is
\be
\CL &=& \frac{1}{2} (\partial_\mu \phi_i)  (\partial^\mu \phi_i) - \frac{1}{2} m^2 \phi_i \phi_i - \frac{\lambda}{4 N} (\phi_i \phi_i)(\phi_j \phi_j).
\ee
The LC truncation approach to this theory was analyzed in \cite{Katz:2016hxp}, to which we refer the reader for more details.  Here we will briefly summarize the conformal primary basis states and Hamiltonian matrix elements in this basis.  As in the CS theory, the basis states for the $\CO(N)$ singlet sector at infinite $N$ are two-particle currents $|\ell, s,\mu\>$, parameterized by a spin $\ell$, parity $s$, and momentum-squared $\mu^2$.
 The interaction preserves parity, so we can focus entirely on the parity-even states and drop the $s$ label for conciseness.

The kinetic term is the same as (\ref{eq:kinterm}),
\begin{equation}
\mathcal M_{\ell',\,\ell}^{\rm kin}
	= 2\pi \delta(\mu^2 - \mu'^2)\delta_{\ell \ell'} \cdot \mu^2\, ,
\end{equation}
and the mass term in the Dirichlet basis is the same as (\ref{eq:MmassDirichlet}):
\begin{equation}
\mathcal M^{\rm mass}_{\ell',\, \ell}
	= 2\pi \delta(\mu^2 - \mu'^2)\delta_{\ell \ell' {\rm mod} 2} \cdot M_{\ell \ell'}, 
	\quad
M_{\ell\ell'}
	= \frac{8}{3} m^2 \min(\ell,\ell') \sqrt{\frac{\min(\ell,\ell')^2-1}{\max(\ell,\ell')^2-1}}\, .
\end{equation}
Consequently, the quadratic Hamiltonian eigenvalues are the same as in (\ref{eq_eval_sodd_peven}):
\begin{equation}
\mathcal M_\alpha = 
	%\equiv 
	\mu^2 + 4m^2 \sec^2 \phi_{\alpha}\, , \qquad  \phi_{\alpha} \equiv  \frac{\pi \left(\alpha-\frac{1}{2}\right)}{\ell_{\rm max}+1}.
\end{equation}
Here, $\alpha$ denotes the eigenvector number, and runs over $\alpha=1, \dots, \frac{\ell_{\rm max}}{2}$.
Finally, the interaction is
\be
 \mathcal M^{\rm int}_{\ell,\ell'} &=& \frac{\lambda}{2
 \sqrt{(\ell^2-1)(\ell'^2-1)}} \frac{1}{\sqrt{\mu \mu'}}
\ee
The values of $\ell, \ell'$ run over positive even integers, $\ell = 2,4,6, \dots$.

It is much easier to diagonalize the full Hamiltonian if we first change to the basis of eigenstates $|\alpha\>$ of the mass term.  This change of basis can be written as
\begin{equation}
|\alpha,\mu\rangle = \sum_{j=1}^{\frac{\ell_{\rm max}}{2}} e^{(\alpha)}_{j} | 2j,\mu \rangle\,, \qquad
\langle \alpha,\mu|\alpha',\mu'\rangle
	= 2\pi \delta(\mu^2 - \mu'^2) \delta_{\alpha\alpha'}
\end{equation}
for some $e^{(\alpha)}_j$; we will not need the exact form. 
Instead, we will just need the interaction matrix elements in the mass eigenbasis.  These matrix elements turn out to be much simpler than the change of basis itself:
\be
\mathcal M^{\rm int}_{\alpha,\alpha'}&=& \frac{\lambda}{2
} v_\alpha(\mu) v_{\alpha'}(\mu') , \quad v_\alpha(\mu) = \frac{1}{\sqrt{\mu}}\frac{1}{\sqrt{\ell_{\rm max}+1}} \ \ \forall \ \alpha .
\ee
In other words,  $M^2_{\rm int}$ is proportional to the outer product $v^T v$, where $v= \frac{1}{\sqrt{\mu}}\frac{1}{\sqrt{\ell_{\rm max}+1}}(1,1,\dots,1 )$ is just the wavefunction of the singlet operator $\phi_i^2$ in the mass eigenbasis. So, the equation for the eigenstates of the interacting Hamiltonian takes the form
\begin{equation}
\label{eq:ONHamEq}
(q^2 - \mu^2 - 4m^2 \sec^2 \phi_{\alpha} )\delta_{\alpha \alpha'} \psi_{\alpha'}(\mu;q) = \frac{\lambda}{2
} v_\alpha(\mu) \int \frac{d\mu'^2}{2\pi} v_{\alpha'}(\mu') \psi_{\alpha'}(\mu';q) 
\end{equation}
 in the mass basis, where $\psi_\alpha(\mu;q) = \< \alpha, \mu | \psi; q\>$ is the component of the eigenstate $\psi$ in the direction of the $|\alpha\>$ mass eigenstate.  As in the previous section, $q^2$ is the eigenvalue of the mass-squared operator $2P_- H$.  In the free theory, we did not need a separate $\mu^2$ and $q^2$ label on the eigenstates since they were related by $q^2= \mu^2 + 4 m^2 \sec^2 \phi_\gamma$,\footnote{More precisely, the states (\ref{eq_evec_sodd_peven}) of definite $\mu^2$ are also eigenstates of $q^2$, so the wavefunctions were $\psi_\alpha(\mu; q) \propto \delta(q^2 -\mu^2 - 4 m^2 \sec^2 \phi_\alpha)$.} but due to the interaction term the $q^2$ eigenstate wavefunctions will have nontrivial $\mu^2$ dependence.  

\subsection{Optimal Basis}
 
 The key features of this integral equation are that the interaction term is factorizable and that the mass term has been diagonalized. The interaction term depends on $\psi$ only through the definite integral
\be
 \int d\mu'^2 v_{\alpha'}(\mu') \psi_{\alpha'}(\mu';q),
\ee
so the Hamiltonian equation takes the form
\be
(q^2 - \mu^2 - 4m^2 \sec^2 \phi_{\alpha} ) \delta_{\alpha \alpha'} \psi_{\alpha'}(\mu;q) \propto  v_\alpha(\mu) .
\ee
One can therefore immediately write down a complete basis for the solutions to $\psi_\alpha(\mu;q)$:
\begin{equation}
\psi_\alpha(\mu;q) = C_\alpha 2\pi\delta(\mu^2(q,\alpha) - \mu^2 )+S\,  P.V. \frac{v_\alpha(\mu)}{\mu^2(q,\alpha) - \mu^2  },
\end{equation}
where ``P.V.'' denotes ``principal value,'' and as in the previous section $\mu^2(q,\alpha) \equiv q^2 -4 m^2 \sec^2 \phi_\alpha$.   The parameters $C_\alpha, S$ implicitly depend on $q^2$, but not $\mu^2$.  

Substituting this general form back into the Hamiltonian equation of motion, the problem is converted from an integral equation to a simple linear equation for the $C_\alpha, S$ coefficients:
\be
\label{eq:ONSeq}
S &=& \frac{\lambda}{2
} \sum_\alpha \Big( C_{\alpha} v_{\alpha}(\llb \mu^2(q,\alpha)\rrb^{1/2})  - \frac12 S \, v^2_\alpha ( \llb -\mu^2(q,\alpha)\rrb^{1/2} )\Big),
\ee
where $v_\alpha(\llb x \rrb^\gamma)$ denotes $v_\alpha(x^\gamma) \Theta(x)$, and we have used the following identity:
\be
\int_0^\infty  d \mu^2 v^2_\alpha(\mu)  P.V. \frac{1}{x-\mu^2} =- \pi v^2_\alpha(\sqrt{-x}) \Theta(-x) \equiv - \pi v^2_\alpha(\llb -x \rrb^{1/2}).
\ee
The equation (\ref{eq:ONSeq}) should be thought of as an equation for $S$ in terms of the parameters $C_\alpha$, because there are actually $\ell_{\rm max}/2$ solutions here, one for each parameter $C_\alpha$.  The reason there is a large degeneracy of eigenvalues is that  $\mu$ is a continuous parameter, so for any value of $q$ we have one eigenstate for each $\alpha$ such that $q^2 > 4 m^2 \sec^2 \phi_\alpha$.  However, it is also clear from (\ref{eq:ONSeq}) that we can separate this space of solutions into one direction parallel to the vector $v_\alpha(\llb \mu^2(q,\alpha) \rrb^{1/2})$, and the space perpendicular to this vector.  The perpendicular space has the trivial solution $S=0$, since it does not see the interaction term.   In other words, we can solve the Hamiltonian equation of motion because for any $\ell_{\rm max}$, there is really only one state that is affected by the interaction.  The upshot of this discussion is that we can focus on the special case
\be
C_\alpha = C \, v_\alpha(\llb \mu^2(\alpha,q)\rrb^{1/2}).
\ee

We also need to compute the norm of the state $|\psi\>$.  The states are $\delta$ function normalized: 
\be
\< \psi ;q | \psi; q'\> &=& \CN_\psi 2\pi\delta(q^2-q'^2).
\ee
We discuss how to compute these norms in appendix \ref{app:WvNm}; the result is
\be
 \CN_\psi = \left( C^2  + \frac14 S^2  \right) \sum_\alpha \Big(  v^2_{\alpha}(\llb \mu^2(q,\alpha)\rrb^{1/2})\Big).
\ee

Lastly, to obtain the spectral function of the singlet operator $\phi^2 \equiv  \phi_i \phi_i$, we need its overlap with the basis states.  Since the interaction is $(\phi^2)^2$, we have essentially already computed the overlap of $\phi^2$ with the mass eigenstates when we computed the interaction matrix elements.  The overlap is\footnote{More explicitly, the overlap with the Dirichlet states $|\ell,\mu \>$ were computed in \cite{Katz:2016hxp} to be $
\< \phi^2 | \mu, \ell \> =  \frac{1}{\sqrt{\mu}} \frac{1}{\sqrt{\ell^2-1}}$,
and from this one can perform the change of basis to obtain the overlap with the mass eigenstates.}
\be
\< \phi^2 | \alpha, \mu \> =  v_\alpha(\mu) .
\ee
Therefore, $| \phi^2\>$ only has overlap with the eigenstate $|\psi\>$ that is parallel to $v$.  The spectral function reduces to the overlap with this state divided by its norm:
\be
\pi
\rho_{\phi^2}(q) &=&\frac{1}{\CN_\psi} \< \phi^2| \psi;q \> \< \psi;q | \phi^2\>  = \frac{\CP_+(q)}{(1+\frac{\lambda}{4} \CP_-(q))^2+ \frac{\lambda^2 }{16} \CP^2_+(q)},
\label{eq:PhiSqSF1}
\ee
 where we have defined
\be
\CP_\pm(q) &\equiv& \sum_\alpha v_\alpha^2(\llb \pm \mu^2(q,\alpha) \rrb^{1/2}).
\ee
In the limit $\ell_{\rm max}\rightarrow \infty$, the $\CP$ functions simplify to
\be
\CP_+(q) &=&  \int_0^{\pi^{-1} \cos^{-1}(\frac{2m}{q})} \frac{dx}{(q^2- 4 m^2 \sec^2(\pi x))^{1/2}} =  \frac{1}{2q} , \nn\\
\CP_-(q) &=& \int_{\pi^{-1} \cos^{-1}(\frac{2m}{q})}^{\frac{1}{2}} \frac{dx}{( 4 m^2 \sec^2(\pi x)-q^2)^{1/2}} = \frac{1}{\pi q} \coth^{-1} \frac{q}{2m} , \nn\\
\ee
for $q>2m$, and vanish for $q<2m$.  Finally, the spectral function reduces to
\be
\label{eq:onsdexact}
\pi\rho_{\phi^2}(q) &=& \frac{\frac{1}{2q}}{\left(1+ \frac{\lambda}{8 q \pi} \log \left( \frac{q+2m}{q-2m} \right) \right)^2 + \left(\frac{\lambda}{8 q} \right)^2} \theta(q-2m),
\ee
which matches  the known result from the standard covariant large $N$ solution.\footnote{See e.g. \cite{Katz:2016hxp} equation (2.34).}

\subsection{Interaction Subspace Notation}
\label{sec:BraKetNotation}

The large $N$ $O(N)$ model is simple enough that  writing all the states out in components was  sufficiently concise that the resulting equations were still reasonably compact and readable.  However, as the size of the interaction subspace increases, keeping track of all components quickly becomes more of a distraction.  In this subsection, we will rewrite the Hamiltonian equations in Dirac notation, which will make the equations more compact and the fact that we are essentially dealing with a single-state system more transparent.  

We begin by defining the following state $|\phi^2\>$ associated with the operator $\phi^2$, whose wavefunction is
\begin{equation}
\< \alpha, \mu | \phi^2 \> \equiv v_\alpha(\mu), 
\end{equation}
We also define the matrices 
\be
\label{eq:DMatrices}
\hat{D}_{\alpha \alpha'} = 2\pi \delta_{\alpha \alpha'} \delta(\mu^2(q,\alpha)-\mu^2) , \qquad \tilde{D}_{\alpha \alpha'} = P.V. \frac{1}{\mu^2(q,\alpha) - \mu^2}
\ee
to be diagonal and have, respectively, $\delta$ functions or principal value poles on their diagonal, as written above.
 More explicitly,
\begin{equation}
 \< \alpha, \mu| \hat{D} | \phi^2\> \equiv v_\alpha(\mu) 2\pi \delta(\mu^2(q,\alpha)-\mu^2), \quad \< \alpha, \mu| \tilde{D} |\phi^2\> = P.V. \frac{v_\alpha(\mu)}{\mu^2(q,\alpha) - \mu^2}  .
 \end{equation}
  Essentially, $\tilde{D}$ and $\hat{D}$ are just the real and imaginary parts of $(q_+-  H_2)^{-1}$, where $H_2$ is the quadratic part of the Hamiltonian:
  \be
  \frac{1}{2(q_+ - H_2+ i \epsilon)} =  \frac{1}{q^2 - 2  H_2+ i \epsilon} \cong \tilde{D} - \frac{i}{2} \hat{D}.
  \ee

The overlap between states is defined as
\be
\< \chi | \phi\> &\equiv&\sum_\alpha  \int \frac{d\mu^2}{2\pi} \chi_\alpha^*(\mu) \phi_\alpha(\mu).
\ee
Finally, the eigenstates $\psi$ will just be denoted $|\psi\>$, 
\be
\< \alpha , \mu | \psi\> \equiv \psi_\alpha(\mu;q),
\ee
and we will also introduce $|\psi\>^{(0)}$ for the purely $\delta$ function piece of $\psi$:
\be
\< \alpha , \mu | \psi\>^{(0)} \equiv C_\alpha 2\pi \delta(\mu^2(q,\alpha) - \mu^2).
\ee
Now that we have introduced a formalism tailored to the problem, solving the Hamiltonian problem will be fairly quick.  The Hamiltonian equation can be written in the following compact form:
\be
(q^2 \mathbbm{1}-2H_2) |\psi\> = \frac{\lambda}{2} |\phi^2\> \< \phi^2| \psi\>.
\ee
For $\mu^2\ne \mu^2(\alpha,q)$, the inverse of $(q^2-2H_2)$ is $\tilde{D}$, so we can multiply through by $\tilde{D}$ as long as we allow an extra purely $\delta$ function piece $|\psi\>^{(0)}$ whose components are all proportional to $\delta(\mu^2-\mu^2(\alpha,q))$:
\be
|\psi\> =  \frac{\lambda}{2} \tilde{D} | \phi^2\> \< \phi^2 |  \psi\> + | \psi\>^{(0)}. 
\ee
Clearly, $\psi$ can be purely its $\delta$ function piece $|\psi\>^{(0)}$ as long as it is orthogonal to $\< \phi^2 |$.  Therefore the subspace of states with $|\psi\>^{(0)}$ orthogonal to $\< \phi^2|$ is trivial to solve, and we can restrict our attention to the state with $|\psi\>^{(0)} \propto \hat{D} |\phi^2\>$. Acting on this state, the interaction manifestly generates only states of the form
\be
|\psi\> &=& C \hat{D} |\phi^2\> + S \tilde{D} |\phi^2\>.
\ee
The equation for $|\psi\>$ becomes the following equation for $C,S$:
\be 
S =\frac{\lambda}{2} \left(  C \< \phi^2 |\hat{D}| \phi^2\> + S \< \phi^2 |\tilde{D}| \phi^2\> \right).
\ee
The norm of the states is
\be
\CN_\psi &=& (C^2 + \frac{1}{4} S^2) \< \phi^2| \hat{D} | \phi^2 \>.
\ee
It is straightforward to verify that the inner products are given by
\begin{subequations}
\begin{align}
& \< \phi^2|\hat{D}| \phi^2\> =
\CP_+(q), \\
& \< \phi^2 |\tilde{D} | \phi^2 \> = -\frac{1}{2} \CP_-(q).
\end{align} 
\end{subequations}
The spectral function is again obtained from taking the overlaps and dividing by the norm:
\be
\pi \rho_{\phi^2}(q) &=& \frac{| \< \phi^2 | \psi\> |^2}{\CN_\psi}  = \frac{\< \phi^2|\hat{D}| \phi^2\>}{(1- \frac{\lambda}{2} \< \phi^2 |\tilde{D} | \phi^2 \>)^2 +\left( \frac{\lambda \< \phi^2|\hat{D} | \phi^2\>}{4} \right)^2},
\ee
which reproduces (\ref{eq:PhiSqSF1}).

%######################################################################%
%======================================================================%
%======================================================================%
%======================================================================%
%######################################################################%
\section{CS at Infinite $N_f$}\label{sec_CS_Nf}

%======================================================================%
%======================================================================%
%======================================================================%
\subsection{Hamiltonian Eigenstates}

The full Hamiltonian, given in \eqref{eq:fullham}, simplifies at infinite $N_f$. First, the matrix elements of the six-body interaction term $H_6$ vanish, and only `planar' contractions of the four-body term $H_4$ survive. In section \ref{sec:MassDiagonalization}, we worked out the conformal primary states as well as the mass eigenstates of the quadratic Hamiltonian $H_2$. Recall that the mass eigenstates came in two families, a parity even set of states $|\alpha\>$ with $\alpha = 1 ,\dots , \frac{\ell_{\rm max}+1}{2}$, and a parity odd set of states $|\beta\>$ with $\beta = 1 , \dots, \frac{\ell_{\rm max}-1}{2}$ (we restrict to odd $\ell_{\rm max}$).   We will work with the two-particle eigenstates $\psi$ in the mass eigenbasis, so $\psi$ has components $\psi^{(+)}_\alpha(\mu;q)$ in the parity even space as well as components $\psi^{(-)}_\beta(\mu;q)$ in the parity-odd space.  The Hamiltonian equation for the eigenstates takes a similar form to what we saw in (\ref{eq:ONHamEq}) for the $O(N)$ model.  Now, we have
\begin{subequations}
\begin{align}
& (q^2 - \mu^2 - 4 m_f^2 \sec^2 \phi_{\alpha+}) \delta_{\alpha \alpha'} \psi^{(+)}_{\alpha'}(\mu;q)\nn\\
& \qquad = \int d\mu'^2 M^{\rm int}_{\alpha \alpha'}(\mu, \mu') \psi^{(+)}_{\alpha'}(\mu';q) +\int d\mu'^2 M^{\rm int}_{\alpha \beta'}(\mu, \mu') \psi^{(-)}_{\beta'}(\mu';q), \\
& (q^2 - \mu^2 - 4 m_f^2 \sec^2 \phi_{\beta-}) \delta_{\beta \beta'} \psi^{(-)}_{\beta'}(\mu;q) \nn\\
& \qquad= \int d\mu'^2 M^{\rm int}_{\beta \alpha'}(\mu, \mu') \psi^{(+)}_{\alpha'}(\mu';q) +\int d\mu'^2 M^{\rm int}_{\beta \beta'}(\mu, \mu') \psi^{(-)}_{\beta'}(\mu';q) .
\end{align}
\end{subequations}

From the structure of the Lagrangian (\ref{eq:LagNoA}) with $a_\mu$ integrated out, 
\begin{equation}
\mathcal L
	= \bar\Psi (i\cancel \d + m_f) \Psi + \frac{2\pi}{k}j_f^\perp \frac{1}{\d_-} j_f^+ + m_a  \left( \frac{2 \pi}{k} \right)^2 j_f^+ \frac{1}{\partial_-^2} j_f^+\, .
\end{equation}
one might expect that the interaction Hamiltonian in the infinite $N_f$ limit should reduce to something with components only in a two-dimensional subspace, corresponding to the $+$ and $\perp$ components of the current operator. This expectation is correct.  It is particularly clear if we take the limit $m_f=0$ from the very beginning, so that we do not need to use the Dirichlet states.  In this case, the interaction piece of the Hamiltonian is nonzero only on the two independent $\ell=1$ primary states; we perform the explicit analysis of this case in appendix \ref{app:NonDirichlet}.  More generally, in appendix \ref{app:InteractingCSMatrixElements},  we work out the form of the matrix elements $M^{\rm int}$ of the two-particle mass eigenstates at $m_f \ne 0$.   In the bra and ket notation of subsection \ref{sec:BraKetNotation}, we can write the Hamiltonian equation for the eigenstates $|\psi\>$ simply as
\be
(q^2\mathbbm{1}-2 H_2)|\psi\> = \Big[ 2\pi i \lambda \bigl(| j_- \>\< j_\perp|   -  | j_\perp \>  \< j_- | \bigr)  - 8 \pi^2\lambda^2m_a  |j_- \> \< j_-|  \Big] | \psi \> ,
 \ee
or equivalently,
\be
\label{eq:CSHamBraKet}
|\psi\> = \tilde{D} \Big[  2\pi i\lambda \big( | j_- \>\< j_\perp|   -  | j_\perp \>  \< j_- | \big)  - 8\pi^2\lambda^2m_a |j_- \> \< j_-|  \Big] | \psi \> + |\psi\>^{(0)}.
 \ee
Recall that, as in subsection \ref{sec:BraKetNotation}, 
we will use the matrices $\tilde{D}$ and $\hat{D}$ to denote the diagonal matrices with the principal value pole $(\mu^2(\gamma,q)-\mu^2)^{-1}$ or $\delta(\mu^2(\gamma,q)-\mu^2)$ function on their diagonal, respectively. The states $|j_\perp\>$ and $|j_-\>$ are the states corresponding to the current operator in the absence of the gauge boson mass term; their components in the Dirichlet and mass eigenbasis were given in section \ref{sec:FreeFermionCorrelators}.   $|\psi\>^{(0)}$ represents the purely $\delta$ function pieces of the eigenstate $|\psi\>$.   

The form (\ref{eq:CSHamBraKet}) of the Hamiltonian equation makes it manifest that states in $|\psi\>^{(0)}$ perpendicular to $|J_\perp\>$ and $|J_-\>$ are not affected by the interaction, and the eigenstates in these directions are the free ones.  There are therefore only two nontrivial eigenstates, and without loss of generality we can parameterize them as
\be
\label{eq:PsiGenFormCS}
|\psi\> &=& \sum_{i=-,\perp}C_i \hat{D} | j_i\> +  S_i \tilde{D} | j_i\> ,
\ee
similarly to what we did in the $O(N)$ model.  We can write the Hamiltonian equation even more compactly as
\be
|\psi\> &=& \left( \sum_{i,j=-,\perp} V_{ij} \tilde{D} | j_i\> \< j_j| \right) | \psi\> + |\psi\>^{(0)},
\ee
where $
V_{\perp \perp} = 0 ,  V_{--} = - \lambda^2 m_a$, and $ V_{-\perp} = V_{\perp -}^* = i \lambda$.
Substituting the general form (\ref{eq:PsiGenFormCS}) into the Hamiltonian equation gives the following relation between the $S_i$ and $C_i$ coefficients:
\be
\label{eq:CSCSRelation}
S_i = \sum_{j,k} V_{ij} \left( \< j_j |\hat{D}| j_k\> C_k + \< j_j | \tilde{D} | j_k\> S_k\right).
\ee

Following the general discussion in appendix \ref{app:WvNm}, the norm of the eigenstates is
\begin{equation}
\< \psi;q | \psi'; q'\> = \CN_{\psi,\psi'} 2\pi \delta(q^2-q'^2), \quad \CN_{\psi,\psi'} = \left( C_{\psi, i}^\dagger C_{\psi',j} + \frac{1}{4} S_{\psi,i}^\dagger S_{\psi',j}\right) \< j_i |\hat{D}|  j_j\>.
\end{equation}
The eigenstates are simply the states with $S_i$s given in terms of $C_i$s (or vice versa) using equation (\ref{eq:CSCSRelation}), and orthonormalizing using the above inner product.

\subsection{Computing Correlators}

Next we want to use the expression from the previous subsection for the eigenstates in order to compute spectral functions for local operators.  Two natural correlators to consider are the current two-point function $\<JJ\>$ and the stress tensor two-point function $\< TT\>$.  The latter is computed in appendix \ref{app_spin_even} 
and is independent of the interactions, so it is just a free theory computation.  So in this subsection, we will compute the  former, where we will explicitly see the role of the gauge boson mass counterterm $m_a$ in canceling divergences.

To compute the spectral densities for the currents, we also need to take into account the fact that the current is modified by the gauge boson mass term according to equation (\ref{eq:JWithMa}). We can write the modified current state $|j_i, m_a\>$ in terms of the $m_a=0$ current $|j_i\>$ as
\be
|j_i, m_a\> = \sum_j X_{ij} |j_j\>, \qquad X = \left( \begin{array}{cc} 1 & 0 \\ -4\pi i \lambda m_a & 1 \end{array} \right).
\ee
Note that by using equations (\ref{eq:PsiGenFormCS}) and (\ref{eq:CSCSRelation}), we have $\< j_i | \psi\> 
= \sum_j V^{-1}_{ij} S_j$,
and therefore
\be
\label{eq:JMaOverlap}
\<j_i, m_a| \psi\> &=& \sum_j (V_0^{-1})_{ij} S_j, \qquad V_0 \equiv V(X^*)^{-1} = \left( \begin{array}{cc} 0 & 2\pi i\lambda \\ -2\pi i \lambda & 0 \end{array} \right).
\ee
 Finally, the spectral function $\rho_{ij}(q)$ is given by summing over a basis of the  eigenstates:
\be
\pi \rho_{ij}(q) = \sum_{\psi, \psi'} \< j_i, m_a| \psi\> (\CN^{-1})_{\psi,\psi'} \< \psi' | j_j, m_a\>.
\ee
The space of eigenstates is parameterized by the coefficients $S_i, C_i$.  Because equation (\ref{eq:CSCSRelation}) relates them to each other, we can use either the $S_i$s or the $C_i$s to parameterize the physical two-dimensional space.  For instance, we can choose our basis states to be $S_-=0, S_\perp=1$ and $S_-=1, S_\perp=0$, and solve for $C_i$ in terms of $S_i$.  Because of (\ref{eq:JMaOverlap}), using the $S_i$ parameters is particularly convenient. As long as we impose  (\ref{eq:CSCSRelation}) and divide by the inverse Gram matrix $\CN_{\psi, \psi'}$, the spectral function will be independent of the specific basis we choose for doing the sum. The result is
\begin{equation}
\label{eq:FinalSF}
\pi \rho_{ij}(q) = \left( \left( X- \< j | \tilde{D}| j \> V_0\right)^\dagger \< j | \hat{D} |j \>^{-1} \left( X - \< j | \tilde{D}|j  \> V_0 \right) + \frac{1}{4} V_0^\dagger \< j | \hat{D} | j \> V_0 \right)^{-1}_{ij}.
\end{equation}
 To go farther, we need expressions for the matrices $\< j | \hat{D} | j \>$ and $\< j |\tilde{D}| j \>$.  The former are in fact just the free theory spectral function components, and were already computed in (\ref{eq_spec_dens_sim2}):
\be
\label{eq:DhatVsRho}
\< j_i |\hat{D}| j_j\> = \pi \rho_{ij}^{\rm free}(q)  . 
\ee
For instance, $\< j_- | \hat{D} | j_-\> = \frac{1}{\ell_{\rm max}+1}  \sum_\alpha \llb \mu_+^2(\alpha, q)\rrb^{-1/2} \cos^2 \phi_{\alpha+} \cdot \left( \frac{1}{4} \right)$ (recall we are using boosts to set $q_-=1$). This fact is consistent with the above equation for $\rho_{ij}$, since $V_0$ vanishes and $X=1$ in the free theory, where there is no gauge boson counterterm needed.

So the only new matrix elements we need to compute are $\< j_i | \tilde{D}| j_j\>$, which  are of the form
\be 
\< j_i | \tilde{D} | j_j\> = \sum_\gamma \int \frac{d\mu^2}{2\pi} P.V. \frac{\< j_i | \gamma, \mu\> \< \mu; \gamma | j_j\>}{\mu_\pm^2(\gamma,q) - \mu^2 }.
\ee
The overlaps of the currents with the parity-even Hamiltonian eigenstates decay like $\mu^{-1/2}$, and consequently they lead to convergent integrals in $\< j_i | \tilde{D} | j_j\>$, of the form
\be
\int \frac{d\mu^2}{2\pi \mu} P.V. \frac{1}{x-\mu^2} = - \frac{1}{2} \llb -x \rrb^{-\frac{1}{2}}.
\ee
  However, the overlaps with the parity-odd eigenstates grow like $\mu^{1/2}$.  This growth  leads to divergences in $\< j_\perp | \tilde{D} | j_\perp\>$.  If we take a hard cut-off on $\Lambda$, then the divergent integrals take the form
\be
\int_{0 \le \mu \le \Lambda} \frac{d\mu^2}{2\pi}  P.V. \frac{\mu}{x-\mu^2} = -\frac{ \Lambda}{\pi} + \frac{1}{2} \llb -x \rrb^{\frac{1}{2}} + \CO\left( \frac{1}{\Lambda} \right).
\ee
By contrast, something like dimensional regularization would simply discard the linear term in $\Lambda$ and keep the finite piece.  For now, we will keep the linear divergence as above.  

Finally, note that the matrix elements $\< j_i | \tilde{D} | j_j\>$ always appear in the spectral function in the combination
\be
X_{ij} - \< j_i | \tilde{D} | j_k\> V_{0,kj} = \delta_{ij} - 2\pi i \lambda \left( \begin{array}{cc} - \< j_- | \tilde{D} | j_\perp \> & \< j_- | \tilde{D} | j_-\> \\
2m_a - \< j_\perp | \tilde{D} | j_\perp \> & \,  \< j_\perp | \tilde{D} | j_-\> \end{array} \right). 
\ee
Crucially, the component $\< j_\perp | \tilde{D} | j_\perp\>$ and the gauge boson mass $m_a$ always come in the combination $2m_a - \< j_\perp | \tilde{D} | j_\perp\>$, which means that the mass term successfully removes the UV divergences as promised!  

\subsubsection*{Case 1: $m_f\rightarrow 0$}

Let us first see how this works in detail in the massless limit $m_f \rightarrow 0$.\footnote{The limit $m_f\rightarrow 0$ is slightly different from taking $m_f=0$ from the beginning, because in the limiting case we are still working within the Dirichlet subspace of states.} The free piece $\rho_{ij}^{\rm free}$ reduces to a diagonal matrix, with the following diagonal components:
\begin{subequations}
\begin{align}
& \pi \rho_{--}^{\rm free}(q) = \frac{1}{\ell_{\rm max}+1} \sum_\alpha \frac{1}{q}  \cos^2 \phi_{\alpha+} \left( \frac{1}{4} \right)  =  \frac{1}{16 q}  , \\
& \pi \rho_{\perp \perp}^{\rm free}(q) =   \frac{1}{\ell_{\rm max}+1} \sum_\beta q  \sin^2 \phi_{\beta-} \left( \frac{1}{4} \right) =  \frac{q(\ell_{\rm max}-1) }{16 (\ell_{\rm max}+1)} .
\end{align}
\end{subequations}

The finite pieces of the matrix elements $\<j_i | \tilde{D} | j_j\>$ all vanish at $m_f\rightarrow 0$, because they are proportional to $\llb -\mu^2(\gamma,q)\rrb^{\pm \frac{1}{2}} = \llb - q \rrb^{\pm \frac{1}{2}} =0$.  The only nonvanishing  piece is the divergence in $\< j_\perp | \tilde{D} | j_\perp \>$:
\begin{equation}\label{eq_jpjp_div}
\< j_\perp | \tilde{D} | j_\perp\> = \sum_\beta   \frac{  -\Lambda/\pi }{\ell_{\rm max}+1} \sin^2 \phi_{\beta -} \left( \frac{1}{4} \right) = \left( \frac{\ell_{\rm max}-1}{\ell_{\rm max}+1}\right) \left( \frac{-\Lambda }{16 \pi} \right).
\end{equation} 
We choose $m_a $ to cancel this piece.  Substituting these expressions into our result (\ref{eq:FinalSF}) for the full spectral function, we find
\be
\lim_{m_f\rightarrow 0} \pi \rho_{ij} = \frac{1}{ (1+ \left( \frac{\pi\lambda }{16}\right)^2 \frac{\ell_{\rm max}-1}{\ell_{\rm max}+1})} \left( \begin{array}{cc} \frac{1}{16 q} & 0 \\ 0 & \frac{q}{16} \end{array} \right).
\ee

\subsubsection*{Case 2: $\ell_{\rm max}\rightarrow \infty$}

Next, we want to compute the spectral function at nonzero $m_f$ and take the truncation limit $\ell_{\rm max} \rightarrow \infty$.  In this case, as we explicitly saw in section \ref{sec:FreeFermionCorrelators}, the free theory spectral function reduces to
\be
\pi \rho_{ij}^{\rm free}(q) =\left( \begin{array}{cc} \frac{4m_f^2+ q^2}{16 q^3} &  - i \frac{m_f}{4 q} \\
  i \frac{m_f}{4 q} & \frac{4m_f^2+ q^2}{16 q} \end{array}\right)_{ij} \theta\left( \frac{q}{m}-2\right).
 \ee
To obtain the interacting spectral function from our expression (\ref{eq:FinalSF}), we also need to compute the matrix elements $\< j_i | \tilde{D} | j_j\>$ at nonzero $m_f$.  In the $\ell_{\rm max} \rightarrow \infty$ limit, all the sums over $\alpha$ and $\beta$ can be approximated as integrals, which are evaluated in appendix \ref{app:InteractingCSMatrixElements}.  We must also choose the gauge boson mass counterterm $m_a$.  One way to do this is to note that Lorentz- and gauge-invariance constrain the correlator to be of the form in equation (\ref{eq_rho_jj}), and in particular $q^2 \rho_{--}(q) $ and $ \rho_{\perp \perp}(q)$ must be equal for both Lorentz and gauge invariance to hold.  This condition together with the form of the spectral function (\ref{eq:FinalSF}) and the matrix elements (\ref{eq:jjtildeIntegrals}) is sufficient to fix the counterterm:
\be
2 m_a = -\frac{1}{8\pi}  \left( \frac{\Lambda}{2} + (2+4\log 2)m_f\right).
\ee

Taking this choice of $m_a$, we find  
\be
X_{ij}- \< j_i | \tilde{D} | j_k \> V_{0,kj} = \delta_{ij} +\frac{ \lambda}{2} \left( \begin{array}{cc} \textrm{Re}(\kappa_0) & \frac{2\pi i \textrm{Im}(\tau_0)}{q} \\
 - 2\pi i \textrm{Im}(\tau_0) q & \textrm{Re}(\kappa_0) \end{array}\right)_{ij},
 \ee
in terms of the free theory $\kappa_0, \tau_0$ from (\ref{eq:taukappa0}).  Note that $V_0 (X-\< j | \tilde{D} | j\>) = (X-\< j | \tilde{D} | j\>)^\dagger V_0$, which allows us to rewrite our expression (\ref{eq:FinalSF})  for the spectral function in the  form
\begin{equation}
\pi \rho_{ij}(q) = \left( \left( X- (\< j | \tilde{D} | j \> - \frac{i}{2}  \< j | \hat{D} | j \>)V_0 \right)^\dagger \< j | \hat{D} | j \>'^{-1} \left(   X- (\< j | \tilde{D} | j  \> - \frac{i}{2}  \< j | \hat{D} | j \>)V_0 \right) \right)^{-1}.
\end{equation}
 We also have, from the free theory results, that
\be
 \< j_i | \hat{D} | j_j\> = \pi \rho_{ij}^{\rm free}=  \left( \begin{array}{cc} \frac{1}{q} \textrm{Re}(\tau_0)  &\frac{1}{2\pi i} \textrm{Im}(\kappa_0) \\
-\frac{1}{2\pi i} \textrm{Im}(\kappa_0)  & q \textrm{Re}(\tau_0)\end{array} \right)_{ij}.
\ee  
This free theory spectral function is the Hermitian piece $G_0^H$ of the free correlator $G_0$, which also has an anti-Hermitian piece $G_0^A$:
\be
 G_0 &=&  \left( \begin{array}{cc} \frac{1}{q} \tau_0  &- \frac{1}{2\pi } \kappa_0 \\
\frac{1}{2\pi } \kappa_0  & q \tau_0 \end{array} \right) =  G_0^{H} +  G_0^{A}.
\ee 
By inspection, we see that our spectral function $\rho_{ij}$ can be written in terms of $G_0^H$ and $G_0^A$ as
\be
\label{eq:RhoConfSimplified}
 \rho_{ij}(q) &=& (1+\frac{i}{2}  G_0^\dagger V_0 )^{-1} G_0^H ( 1- \frac{i}{2} V_0 G_0 )^{-1}.
\ee

The full correlator $G$ in the interacting theory can be solved by a standard resummation of 1PI diagrams; in this picture, $V_0$ is just the gauge boson propagator at $q_-=1$.  The result is that
\be
\label{eq:CovariantResummationCS}
G = G_0 \left( 1 - \frac{i}{2} V_0 G_0\right)^{-1}.
\ee
 
 The Hermitian piece of this correlator is
\begin{equation}
G^H = \frac{G_0(1-\frac{i}{2} V_0 G_0)^{-1} + (1+\frac{i}{2}  G_0^\dagger V_0)^{-1} G_0^\dagger}{2} 
  = (1+\frac{i}{2}  G_0^\dagger V_0)^{-1} G_0^H (1-\frac{i}{2} V_0 G_0)^{-1},
  	\label{eq:FinalJJ}
\end{equation}
which agrees with our conformal truncation result (\ref{eq:RhoConfSimplified}).

\subsection{Convergence Rate}

The large $N_f$ limit is a rare case where conformal truncation can be evaluated analytically as a function of the truncation level $\ell_{\rm max}$, and more generally the analysis is limited by numerical resources to some finite maximum value. The rate at which the truncated result approaches the exact result as a function of $\ell_{\rm max}$ is an important part of how useful conformal truncation can be in practice.   In this subsection, we will consider the size of the corrections to the exact spectral functions at finite $\ell_{\rm max}$.  

\begin{figure}[t!]
\begin{center}
\includegraphics[width=0.9\textwidth]{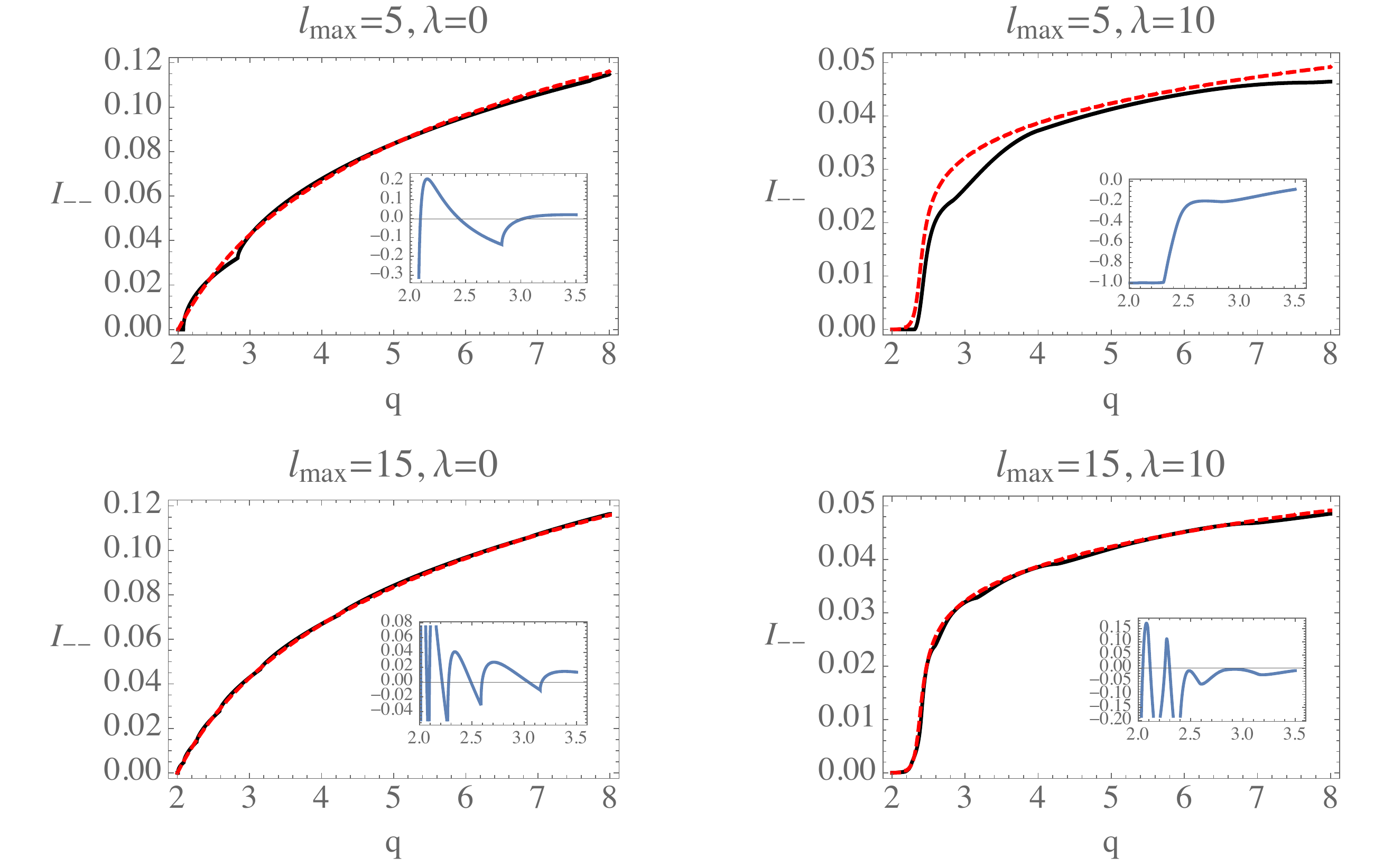}
\caption{Comparison with exact analytic result (red, dashed) and truncation result (black, solid) for various values of the coupling $\lambda$ and truncation level $\ell_{\rm max}$. The residuals (defined as $\frac{\rm truncation}{\rm exact}-1$) are shown in the insets.}
\label{fig:NumericComparisons}
\end{center}
\end{figure}

We must first discuss what quantities we want to compare at finite vs infinite $\ell_{\rm max}$.  It is clear by inspection of the terms in the sums of the spectral functions (\ref{eq_spec_dens_sim2}) that the spectral functions themselves have inverse-square-root  singularities, $\sim \frac{1}{\sqrt{\mu^2-\mu^2(\alpha,q)}}$, at any finite $\ell_{\rm max}$, whereas the exact spectral functions are smooth functions of $q$.  So, the finite $\ell_{\rm max}$ spectral functions do not converge pointwise.  Rather, we must consider the spectral functions integrated against sufficiently smooth kernels.  In practice, taking the integrated spectral functions,
\be
\label{eq:intsd}
 I_{\mu\nu}(q) \equiv \int_0^{q^2} d\mu^2  \rho_{\mu\nu}(\mu),
\ee
will be sufficient.  Comparisons between the exact result for $I_{--}$ and the truncation result at finite $\ell_{\rm max}$ are shown in Fig. \ref{fig:NumericComparisons} for a couple of values of coupling $\lambda$ and $\ell_{\rm max}$. Plots for the other components $I_{\mu\nu}$ are qualitatively similar. 

\begin{figure}[t!]
\begin{center}
\includegraphics[width=0.45\textwidth]{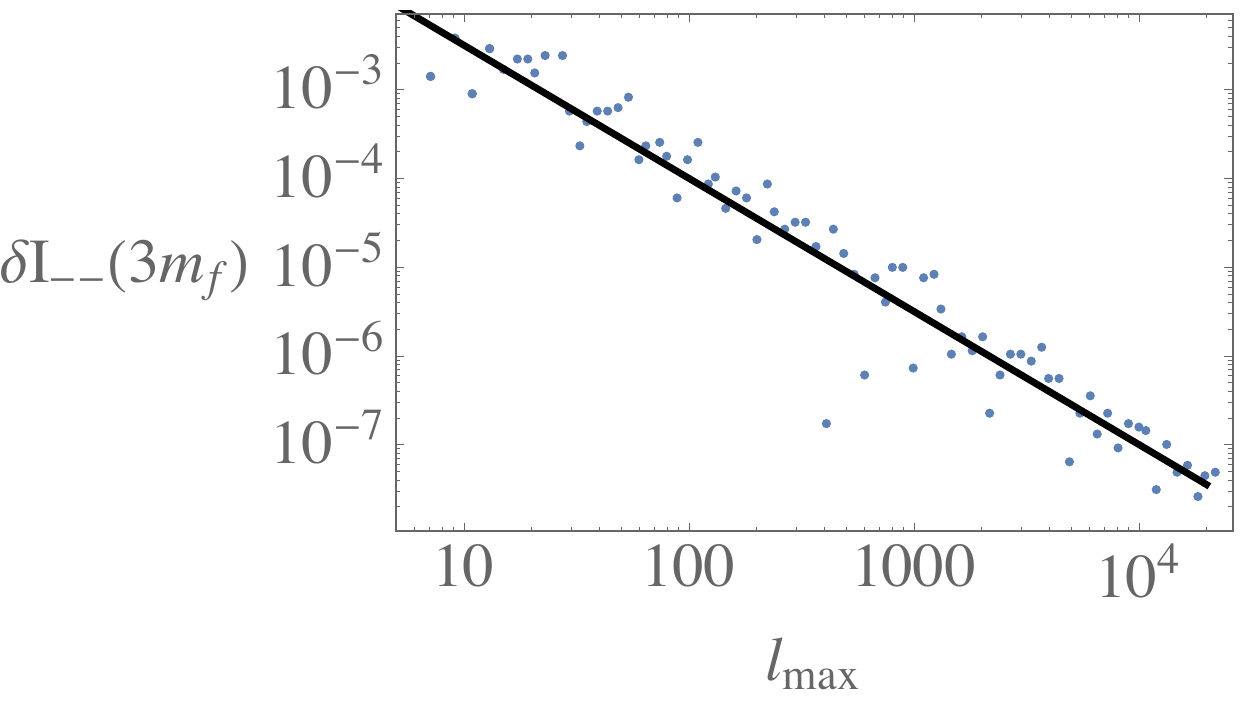}
\includegraphics[width=0.45\textwidth]{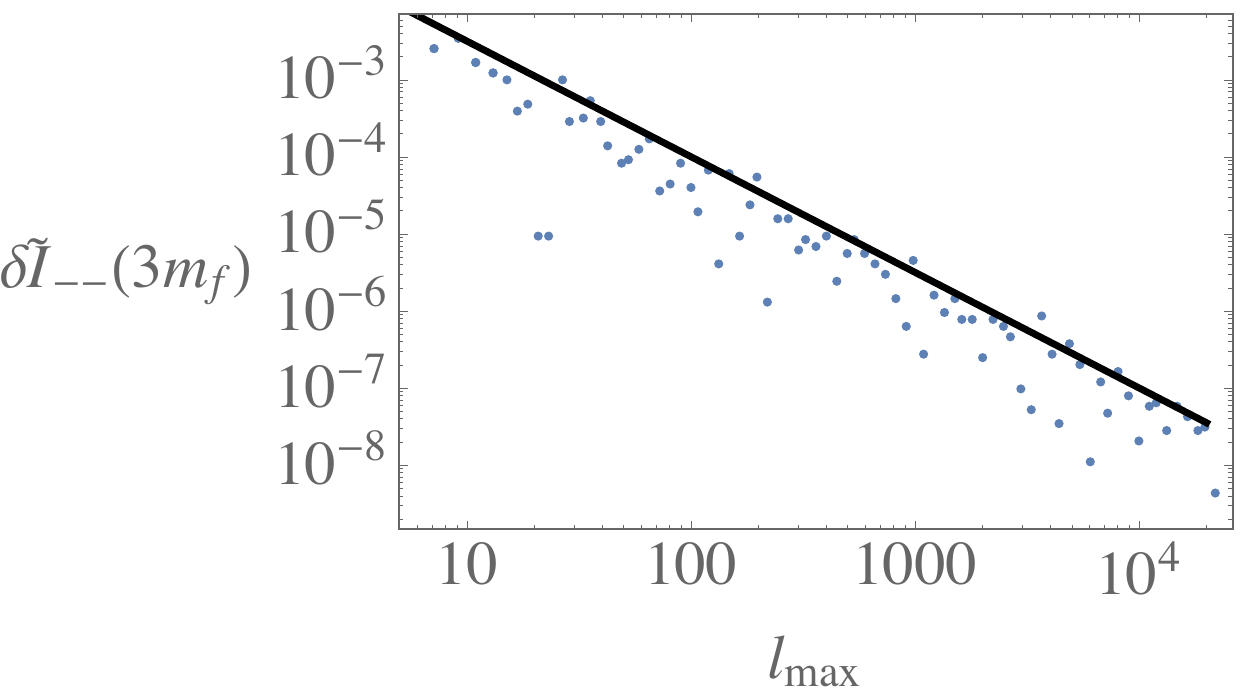}
\caption{{\it Left}: Difference $\delta I_{--}$ between exact analytic result  and truncation result at $\lambda=0$ and $q=3m_f$ as a function of truncation level $\ell_{\rm max}$. The black solid line is 0.1 $\ell_{\rm max}^{-3/2}$ for comparison. {\it Right}: Analogous plot, but for the integrated value of $\< j_- | \tilde{D} | j_-\>$. }
\label{fig:convergencerate}
\end{center}
\end{figure}

We can also read off the rate at which the integrated spectral function from truncation approaches the exact result by looking at the difference as a function of $\ell_{\rm max}$ for some fixed value of $q$.  In Fig.~\ref{fig:convergencerate}, we show the result for $\lambda=0$ and $q=3 m_f$.  To quantify the rate of convergence for the interactions as well, we also look at the difference ``$\delta \tilde{I}_{--}(3m_f)$'' between the integrated value  of the matrix element $\< j_- | \tilde{D}| j_-\>$ from $q=2m_f$ to $q=3m_f$ and its $\ell_{\rm max}=\infty$, analytic value.  For comparison, we show  $\sim \ell_{\rm max}^{-3/2}$, which fits the overall trend. There is some spread around this trendline, with some values of $\ell_{\rm max}$ happening to be better or worse at the particular point $q=3m_f$ we have chosen for the plot. 

Finally, we would like to comment on another quantity related to the spectral density,
which we have found to converge even faster than \eqref{eq:intsd}. For any 
spectral density $\rho$ with a mass gap at $m$ one can define its Fourier transform
\begin{equation}
\label{eq:tilderho}
\tilde{\rho}(x) \equiv \int_{4 m^2}^\infty d \mu^2 e^{i \mu^2 x} \rho(\mu).
\end{equation}
Note that this is essentially the Wightman function as a function of lightcone time $x^+$ at finite $q_-$, i.e. $G(x^+, q_-, q_\perp=0) \sim \int d\mu^2 e^{-i \frac{\mu^2}{2q_-} x^+} \rho(\mu).$  For simplicity, we will focus on the case of the $O(N)$ model, 
whose exact spectral density at finite 
$\ell_{\rm max}$ was shown in equation \eqref{eq:PhiSqSF1}.

\begin{figure}[t!]
\begin{center}
\includegraphics[width=0.45\textwidth]{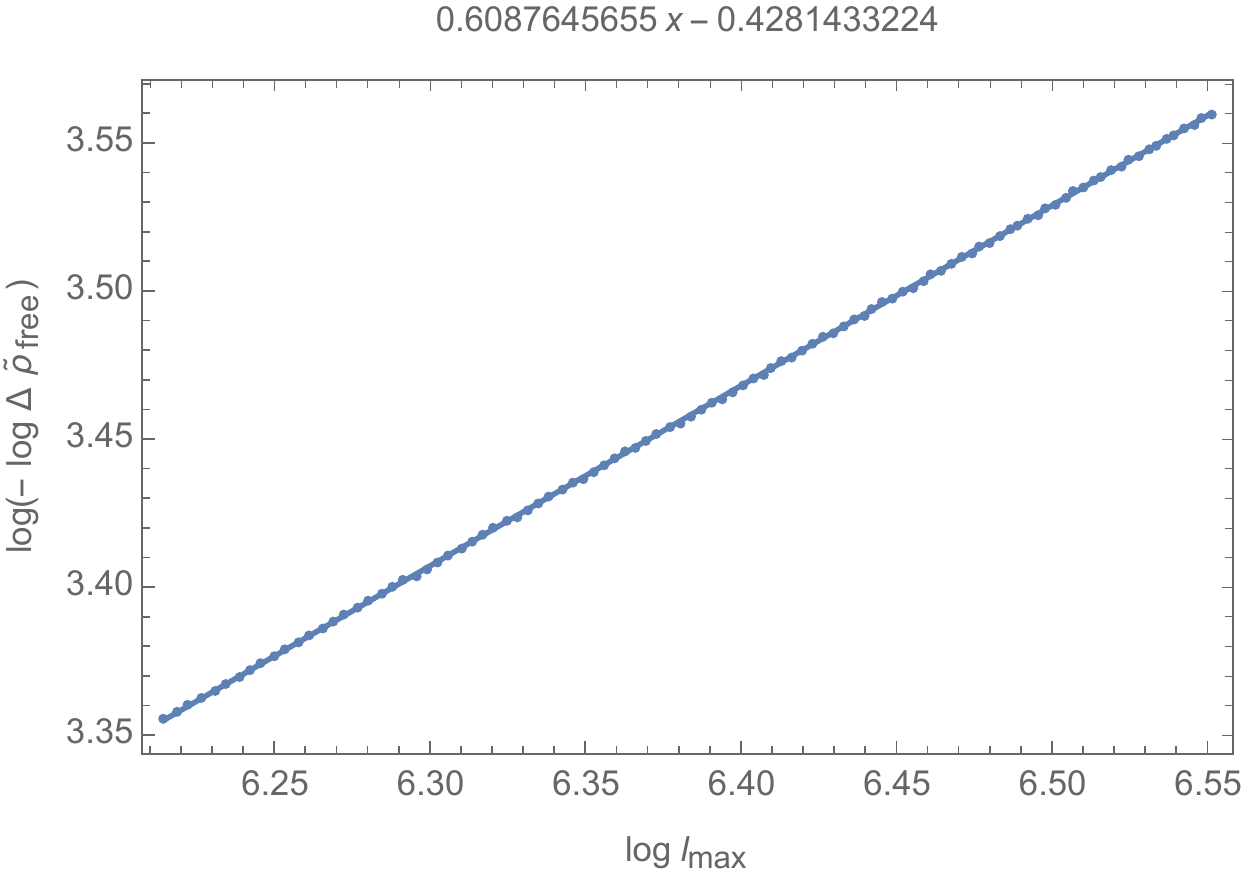}
\includegraphics[width=0.45\textwidth]{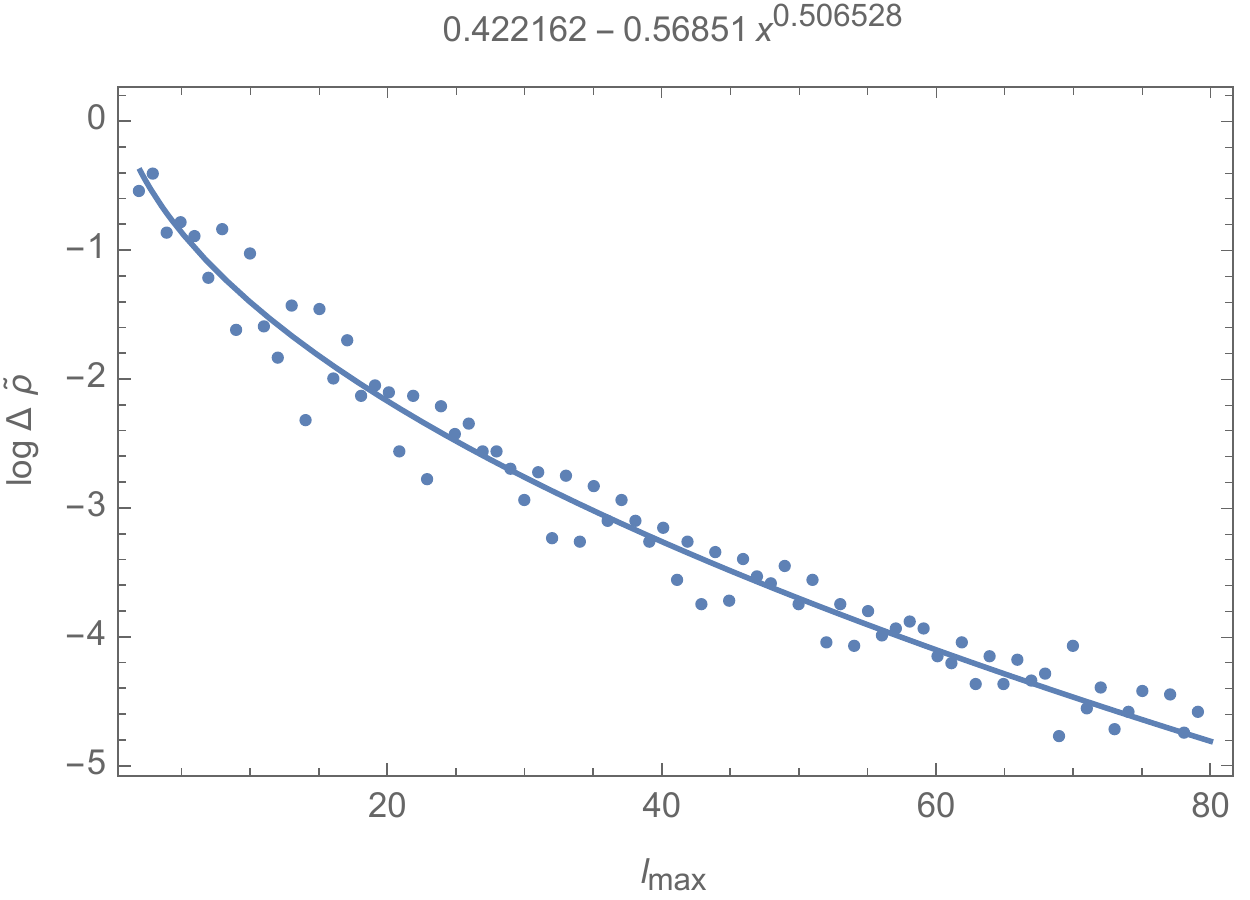}
\caption{
{\it Left}: Dependence of $\log\left( -\log \Delta \tilde{\rho}_{\rm free}\right)$ versus
$\log \ell_{\rm max}$, 
where $\Delta \tilde{\rho} \equiv \frac{\tilde{\rho}-\tilde{\rho}^{\rm num}}
{\tilde{\rho}}$  and $\tilde{\rho}^{\rm num}$ is evaluated at finite $\ell_{\rm max}$,
at $x = 0.2 + 0.01 i$. The linear best fit is shown above the plot. There is a trend 
$\Delta \tilde{\rho} \sim e^{- a L^b}$ with $b \sim 0.6$.
{\it Right}:
Dependence of $\log \Delta \tilde{\rho}$ versus $\ell_{\rm max}$ for $\lambda = 24$ and
$x = 0.16 e^{0.2 i}$. Exponential best fit is shown above the plot.}
\label{fig:rhotilde}
\end{center}
\end{figure}

In figure \ref{fig:rhotilde} is shown the relative error between the numerical and exact $\tilde{\rho}_{\phi^2\phi^2}(x)$ in the free and interacting cases. 
$x$ has to have an imaginary part in order to regulate the UV divergence in 
integral \eqref{eq:tilderho}. In the interacting case, we choose $\lambda$ so that there exist a strongly coupled region $2 m \lesssim q \lesssim
\frac{\lambda}{16}$ (by inspection of \eqref{eq:onsdexact}), 
and correspondigly we choose $x$ to probe the spectral density in this window, 
$x \approx \frac{1}{q^2}$. Empirically, we find that in both cases the convergence is exponentially fast, 
$|\Delta \tilde{\rho}_{\phi^2 \phi^2}| \sim e^{- a \ell_{\rm max}^b}$. 
We find numerically that $b\sim \frac{2}{3}$, however we haven't investigated the 
analytical dependence of $a$ on $x$. We imagine it could be possible to extract 
physical quantities, such as mass gaps and anomalous dimensions, from the $x$ 
dependence of $\tilde{\rho}$. We leave this interesting question for future work.

\section{Discussion and Future Directions}
\label{sec:future}

In this paper, we have attempted to make progress in understanding how Lightcone Conformal Truncation can be applied to gauge theories in $d>2$ dimensions.  We have focused on 3d Chern-Simons theories coupled to matter since, in terms of the degrees of freedom in the gauge field, such theories are a natural next step after successful work in 2d Yang-Mills theories. Unlike in 2d, the interaction in our case is dimensionless, and one might have worried a priori that conformal truncation methods require strictly {\it relevant} interactions in order to converge as the truncation parameter $\Delta_{\rm max}$ is lifted to infinity. Instead, the complications from the increased dimensionality of the interaction term turned out to be manageable.  Specifically, the interaction leads to UV divergences, which must be regulated.  Taking a hard cutoff on the Lorentz-invariant momentum-squared as we did here produces finite answers, but also breaks gauge invariance and generates a gauge boson mass, which can be subtracted off by including a corresponding counterterm in the original Hamiltonian.

Our analysis was restricted to the case of Abelian gauge theories in the large $N_f$ limit.  This restriction was partly for simplicity and clarity - in this limit, the entire conformal truncation analysis can be performed analytically, and compared to the known exact answer - but mostly to avoid some conceptual puzzles about how to implement the method that arise at small $N_f$.  These puzzles, described briefly in section \ref{sec:infinitenf}, must be resolved in order to apply LC Conformal Truncation to the general case of Chern-Simons matter theories.  
 A relatively simple setting where some of those problems
can be isolated is the large $N_c$ limit, where the exact solution is also known from more standard covariant methods \cite{GurAri:2012is}. In this limit, the rainbow diagrams that renormalize the fermion propagator can be computed exactly, and one might hope to gain insight into how to properly define the LC Hamiltonian with the gauge boson and nondynamical fermion component $\chi$ integrated out.
One could also attempt to define the Hamiltonian indirectly via the Bethe-Salpeter equation, as in appendix \ref{sec:bs}.\footnote{A related computation was performed in 
\cite{Jain:2014nza}, where the authors attempted to derive the singlet channel $2\to 2$ S-matrix in large $N_c$ CS theories from a certain kinematical configuration 
of the momentum space correlator. For technical reasons they did not evaluate it directly, but it would be interesting to return to this problem.}
We leave this important step for future work.

From a condensed matter perspective, it would be interesting to try to incorporate  Lorentz violating deformations such as a chemical potential or a background magnetic field. These can in principle be studied within the same framework, which could potentially help address the problem of a Fermi surface coupled to critical bosons. It would also be very interesting to try to generalize the method to nonzero temperature and study finite temperature quantum criticality. Accessing the real time dynamics of strongly correlated systems is notoriously difficult, even numerically.

One shortcoming of our approach is that we use an IR regulator that breaks Lorentz invariance, and Lorentz invariance is restored only when the truncation level $\ell_{\rm max}$ is lifted to infinity.  An IR regulator is needed in order to deal with the matrix elements associated with the fermion mass term, and our regulator effectively treats the lightcone direction $x^-$ differently from the other spacetime directions.  It would be more satisfying, and likely useful, to develop a regulator that preserves Lorentz invariance.

Finally, although the interaction in our analysis is dimensionless, its effect is relatively mild in that it does not change the scaling dimension of any operators.  A more stringent test of the method would be to consider a marginal deformation that allows one to dial anomalous dimensions of the deformed theory, or even to induce log running of couplings.  Applying LC conformal truncation to such examples, even two-dimensional ones, would likely shed light on the details of how and whether the technique works more generally.

\section*{Acknowledgments} 
 
We would like to thank Jeremias Aguilera-Damia and Xi Yin for correspondence, and especially Matthew Walters for many discussions. We would also like to thank Matthew Walters and Max Zimet for collaboration at an early stage of the project.
ALF and EK were supported in part by the US Department of Energy Office of Science under Award Number DE-SC-DE-SC0015845, and ALF was supported in part by a Sloan Foundation Fellowship. ALF, EK, and LGV were also supported in part by the Simons Collaboration Grant on the Non-Perturbative Bootstrap. LD was supported by the US Department of Energy Office of Science under Award Number DE-SC0018134.

\appendix

%######################################################################%
%======================================================================%
%======================================================================%
%======================================================================%
%######################################################################%
\section{Algebra}\label{app_alg}

%======================================================================%
%======================================================================%
%======================================================================%
\subsection{Hamiltonian and mode expansion}\label{app_H_modes}

The Lagrangian \eqref{eq_L_final_ma} leads to the Hamiltonian density
\begin{equation}
\mathcal H 
	= i \sqrt{2} \chi_{\rm os}^*\d_- \chi_{\rm os} + 2m_a \left(\frac{\pi}{k}\right)^2 \psi^*\psi \frac{1}{\d_-^2} \psi^*\psi
\end{equation}
with anti-commutation relation
\begin{equation}
\{\psi(x),\psi^*(y)\}
	= \frac{1}{\sqrt{2}} \delta(x^--y^-)\delta(x^\perp - y^\perp)\, .
\end{equation}
The fields can be expanded into modes as
\begin{equation}\label{eq_modes}
\psi(x) 
	= \frac{1}{2^{1/4}}\int \frac{d^2p}{(2\pi)^2} e^{ip\cdot x}\psi(p)
	= \frac{1}{2^{1/4}}\int \frac{d^2p}{(2\pi)^2} e^{ip\cdot x}\left(a_{-p}\Theta(-p_-) + b^\dagger_p \Theta(p_-)\right)\, .
\end{equation}
The step functions are necessary because there are no modes with $p_-<0$. The creation and annihilation operators satisfy 
\begin{equation}
\{a_p,a_q^\dagger \} =
\{b_p,b_q^\dagger \} 
	= (2\pi)^2\delta(p_--q_-)\delta(p_\perp-q_\perp)\, .
\end{equation}
The Hamiltonian has  two-particle, 4-particle, and six-particle terms:
\be
H &=& H_2 + H_4 +H_6. 
\ee 
Using the notation $\psi^*(-p) = \psi(p)^*$,  in momentum space these take the form
\begin{equation}
\label{eq:fullham}
\begin{split}
H_2 =& 	\int \frac{d^2 p}{4\pi^2} \, h_2\, \psi^*(p)\psi(-p) , \\ 
H_4 =&	\int \frac{d^2 p}{4\pi^2}\frac{d^2 q}{4\pi^2}\frac{d^2 q'}{4\pi^2} \, h_4\, \psi_i^*(-p)\psi_j^*(q-q')\psi_i({p-q}) \psi_j(q'), \\
H_6	=& \int \frac{d^2 p}{4\pi^2}\frac{d^2 q}{4\pi^2}\frac{d^2 p'}{4\pi^2}\frac{d^2 q'}{4\pi^2}\frac{d^2 p''}{4\pi^2} \, h_6\, \psi_i^*(-p-q)\psi_j^*(q-p')\psi_k^*(q'-q'') \psi_i(p-q')\psi_j(p')\psi_k(q''),
\end{split}
\end{equation}
with
\begin{subequations}\label{eq_h2h4h6}
\begin{align}
h_2 \label{eq_h2}
	&= \frac{p_\perp^2 + m_f^2}{2p_-}  \, , \\
h_4 \label{eq_h4}
	&= \frac{\pi}{k} \left[\frac{i(p-q)_\perp - m_f}{(p-q)_- q_-} + \frac{ip_\perp + m_f}{p_- q_-}\right] + 
	m_a \left(\frac{2\pi}{k}\right)^2 
	\frac{1}{q_-^2} \, , \\
h_6
	&= - \left(\frac{\pi}{k}\right)^2 \frac{2}{p_-q_-q'_-}\, .
\end{align}
\end{subequations}
%

%======================================================================%
%======================================================================%
%======================================================================%
\subsection{Higher spin currents and 2-particle states}\label{app_higherspin}

The two-particle primaries in the free fermion theory are spanned by the singlet
\begin{equation}\label{eq_singlet_expl}
J^0 = \bar\Psi \Psi 
	=  \psi^* \chi + \chi^* \psi\, ,
\end{equation}
and the higher spin currents which can be expressed in terms of Gegenbauer polynomials as \cite{Giombi:2017rhm}
\begin{equation}\label{eq_hat_J}
\hat J^\ell(y) = (\hat\d_1 + \hat \d_2)^{\ell-1}C_{\ell-1}^1 \Bigl(\frac{\hat\d_1 - \hat \d_2}{\hat\d_1 + \hat \d_2}\Bigr) :\bar\Psi(y_1) \hat \gamma \Psi(y_2): \Bigr|_{y_{1,2}\to y}\, ,
\end{equation}
where hats denote contractions of tensors with a null vector $\hat A \equiv A_{\mu \nu \cdots}z^\mu z^\nu \cdots $. Current conservation implies that only two components of each current are independent, which we take to be $J^\ell_{-\ldots--}$ and $J^\ell_{-\ldots-\perp}$. We then define the corresponding states as
\begin{subequations}
\begin{align}
|\ell,+,P\rangle
	&\propto \int d^3 y \, e^{-iyP} J^\ell_{-\ldots --}(y) |0\rangle\, , \\
|\ell,-,P\rangle
	&\propto \int d^3 y \, e^{-iyP} J^\ell_{-\ldots -\perp}(y) |0\rangle\, ,
\end{align}
\end{subequations}
with the normalizations to be fixed later. The $\pm$ label denotes the charge under parity. The parity even states can be simply obtained from \eqref{eq_hat_J} by taking $z^\mu = \delta^\mu_-$. The parity odd states can be found for example by taking $z^\mu = \delta^\mu_- + \epsilon \delta^\mu_\perp$ (which is still null to leading order in $\epsilon$), and picking up the $O(\epsilon)$ part of \eqref{eq_hat_J}. Further using the mode expansion \eqref{eq_modes} one finds
\begin{equation}
|\ell,s,P\rangle
	= A_{\ell s}\int_0^1 dx\, \sqrt{x(1-x)} \frac{1}{2} \sum_{\sigma=\pm} f_{\ell s}(x,\sigma) :\psi_p^* \psi_{P-p}: |0\rangle\, ,
\end{equation}
where $x=p_-/P_-$ and $p_\perp = \sigma \mu \sqrt{x(1-x)}$, and the functions are given by
\begin{equation}\label{eq_app_f_ell}
f_{\ell +}(x,\sigma) = 2C_{\ell-1}^{1}(1-2x)\, , \qquad \qquad
f_{\ell -}(x,\sigma) = \frac{\sigma \ell/2}{\sqrt{x(1-x)}}C_{\ell}(1-2x)\, ,
\end{equation}
for $\ell\geq 1$. Here $C_{\ell}(y) \equiv \lim_{m\to 0}\frac{1}{m}C_\ell^m(y)$. Doing the same with the singlet \eqref{eq_singlet_expl} gives 
\begin{equation}\label{eq_app_f_0}
f_{0-}(x) 
	= \frac{\sigma/\sqrt{2}}{\sqrt{x(1-x)}}\, .
\end{equation}
The normalizations of these functions is arbitrary for now, since we have not fixed $A_{\ell s}$. However we will show in Section \ref{app_matrixel} that with the choices \eqref{eq_app_f_ell}, \eqref{eq_app_f_0}, $A_{\ell s}$ will be independent of $\ell$ and $s$.

%======================================================================%
%======================================================================%
%======================================================================%
\subsection{Inner products and matrix elements}\label{app_matrixel}

Two-particle states in the singlet sector take the form
\begin{equation}\label{eq_two_particle_states}
\begin{split}
|\phi,P\rangle
	&= \frac{A_\phi'}{\sqrt{N_f}} \int d^3 y \, e^{-iyP} \left[f_\phi(-i\d^1,-i\d^2) :\psi^*(y_1) \psi(y_2): \right]_{y_{1,2}\to y} | 0 \rangle \\
	&= \frac{A_\phi}{\sqrt{N_f}} \int_0^1 \frac{dx}{\pi} \, \sqrt{x(1-x)} \frac{1}{2}\sum_{p_\perp} f_\phi(p,P-p):\psi^*(p) \psi(P-p): | 0 \rangle 
\end{split}
\end{equation}
where $x\equiv p_-/P_-$, and to get the to the second line we used the mode expansion from \ref{app_H_modes}. The sum is over $p_\perp = \pm \mu \sqrt{x(1-x)}$. We now choose 
\begin{equation}
A_\phi = \sqrt{{P_-}/{\mu}}
\end{equation}
for all states $\phi$. Using Wick contractions, the inner product of two states is given by
\begin{equation}
\langle \phi',P' | \phi, P\rangle
	= 2\pi \delta(\mu^2 - \mu'^2) \int_0^1 \frac{dx}{\pi} \sqrt{x(1-x)} \sum_{p_\perp} f_{\phi'}^*(p,P-p) f_\phi(p,P-p)\, .
\end{equation}
Note that all inner products and matrix elements will contain an overall spatial momentum preserving delta function $(2\pi)^2 \delta^2(\vec P - \vec P')$ which we will not write explicitly. The choices of basis functions \eqref{eq_app_f_ell}, \eqref{eq_app_f_0} is then orthonormal
\begin{equation}\label{eq_orthonormal}
\langle \ell', s',\mu' | \ell, s,\mu\rangle
	= 2\pi \delta(\mu^2 - \mu'^2) \delta_{\ell\ell'} \delta_{ss'}\, .
\end{equation}

Matrix elements of the mass operator
\begin{equation}
\mathcal M \equiv P_\mu P^\mu = 2P_- H\, 
\end{equation}
can be similarly computed. As an example, we can consider a generic two-body hamiltonian
\begin{equation}
\mathcal M_2 = 2P_- \int \frac{d^2p}{4\pi^2} h_2(p) \psi^*(p) \psi(-p)\, .
\end{equation}
Using Wick contractions, one finds that its matrix elements in two-particle states \eqref{eq_two_particle_states} is
\begin{equation}
\begin{split}
&\langle \phi',P' | \mathcal M_2 |\phi, P\rangle = 2\pi \delta(\mu^2 - \mu'^2) \cdot \\
	&\quad  \int_0^1 \frac{dx}{\pi} \sqrt{x(1-x)} \sum_{p_\perp} f_{\phi'}^*(p,P-p) f_\phi(p,P-p)2P_-\left[h_2(-p) - h_2(P-p)\right]\, .
\end{split}
\end{equation}
For example for the Hamiltonian of a free massive particle \eqref{eq_h2} the quantity in brackets is 
\begin{equation}
\mu^2  + \frac{m_f^2}{x(1-x)}\, .
\end{equation}
Since the basis states $|\ell,s,\mu\rangle$ are orthonormal \eqref{eq_orthonormal}, the matrix elements of the kinetic term are just
\begin{equation}
\mathcal M^{\rm kin}_{\ell's',\ell s}
	\equiv \langle \ell', s', \mu'|\mathcal M^{\rm kin}|\ell, s, \mu\rangle
	= 2\pi \delta(\mu^2 - \mu'^2) \delta_{\ell\ell'} \delta_{ss'} \cdot \mu^2\, .
\end{equation}
The matrix elements of the mass term $m_f^2$ have the form
\begin{equation}
\mathcal M_{\ell's',\,\ell s}^{\rm mass}
	\equiv \langle \ell',s', \mu'|\mathcal M^{\rm kin}|\ell,s, \mu\rangle
	= 2\pi \delta(\mu^2 - \mu'^2) \delta_{ss'}\cdot M^s_{\ell \ell'}\, .
\end{equation}
In the parity-even sector ($s=+$) they are given by
\begin{equation}
M_{\ell \ell'}^+ 
	= 8 m_f^2 \int_0^1 \frac{dx}{\pi} \frac{C_{\ell-1}^1(1-2x)C_{\ell'-1}^1(1-2x)}{\sqrt{x(1-x)}}
	= 8 m_f^2 \min(\ell,\ell') \delta_{\ell \ell' \rm mod 2}\,.
\end{equation}
The matrix elements of the mass term in the  parity-odd sector have an IR divergence which is discussed in the next section.

The matrix elements of the interaction terms \eqref{eq:fullham} between two-particle states can be similarly computed. For the quartic interactions $\mathcal M_4 = 2P_-H_4$ we have
\begin{equation}\label{eq_M4_matrixel}
\begin{split}
&\langle \phi',P' | \mathcal M_4 | \phi,P \rangle
	= \frac{P_-^2}{\sqrt{\mu\mu'}}\cdot \\
	&\qquad N_f\int_0^1 \frac{dxdx'}{\pi^2}\sqrt{x(1-x)}\sqrt{x'(1-x')} \frac{1}{4}\sum_{p_\perp ,p_\perp '}f_{\phi'}(p',P-p') f_{\phi}(p,P-p) \tilde h_4\, ,
\end{split}
\end{equation}
with
\begin{equation}\label{eq_h4tilde}
\begin{split}
\tilde h_4 &= 
	-h_4(-p',-P,-p) - h_4(P-p,P,P-p')\\
	&+ \frac{1}{N_f}\left[h_4(P-p,p'-p,-p)+h_4(-p',p-p',P-p')\right]\, ,
\end{split}
\end{equation}
where $h_4$ is given in eq.~\eqref{eq_h2h4h6}. In the $N_f\to \infty$ limit, $H_6$ does not contribute and the second line in \eqref{eq_h4tilde} vanishes. In this limit one has
\begin{equation}
\begin{split}
N_f {P_-^2 }\tilde h_4= & -{{8 \pi^2}\lambda^2 m_a} - \pi \lambda m_f \left[\frac{1}{x(1-x)} + \frac{1}{x'(1-x')}\right] \\
&+ \pi \lambda \left[i\mu \sigma \frac{1-2x}{\sqrt{x(1-x)}} 
- i\mu' \sigma' \frac{1-2x'}{\sqrt{x'(1-x')}} \right]\, .
\end{split}
\end{equation}
When $m_f = 0$, this leads to matrix elements $\mathcal M^{\rm int}_{\ell' s',\ell s}\equiv \langle\ell',s',\mu' | \mathcal M_4 |\ell,s,\mu\rangle$ in the primary basis \eqref{eq_app_f_ell} given by 
\begin{equation}\label{eq_M4_massless_el}
\mathcal M^{\rm int}_{\ell' s',\ell s}
	= - \frac{m_a \pi^2 \lambda^2}{4\sqrt{\mu\mu'}}\delta_{\ell 1}\delta_{\ell'1}\delta_{s+}\delta_{s-}\frac{i\pi\lambda}{8} \sqrt{\frac{\mu}{\mu'}} \delta_{\ell 1}\delta_{\ell' 1} \delta_{s-}\delta_{s'+} + \hbox{h.c.}\, .
\end{equation}
%

%======================================================================%
%======================================================================%
%======================================================================%
\subsection{Projection to Dirichlet basis}\label{app_dirichlet}
The matrix elements of the mass term in the parity-odd sector are given by
\begin{equation}
M_{\ell \ell'}^-
	= \frac{m_f^2\, \ell \ell'}{2^{1+(\delta_{\ell 0}+\delta_{\ell' 0})/2}} \int_0^1 \frac{dx}{\pi} \frac{C_{\ell}(1-2x)C_{\ell'}(1-2x)}{[{x(1-x)}]^{3/2}}\, ,
\end{equation}
and are all divergent. A sharp cutoff on the integral gives
\begin{equation}\label{eq_Mminus}
M_{\ell \ell'}^- = \frac{8  m_f^2 }{2^{(\delta_{\ell 0}+\delta_{\ell' 0})/2}} 
	\left[\frac{1}{\sqrt{\epsilon}} - \max(\ell,\ell')\right] \delta_{\ell \ell'\rm mod 2}\, .
\end{equation}
In the limit $\epsilon\to 0$, two of the eigenvalues of $M^-$ diverge (one in each spin even and odd sectors), and the corresponding states decouple from the theory. It is convenient to remove them by hand by projecting the basis to the kernel of the divergent part of $M^-$. Doing so gives the orthonormal Dirichlet basis
\begin{equation}
f_{\ell -}(x,\sigma)
	= \sigma  \frac{\sqrt{x(1-x)}}{\sqrt{\ell^2-1}}8C_{\ell-2}^2(1-2x)\, .
\end{equation}
In this basis, the matrix elements of the mass term are finite and given by
\begin{equation}
\begin{split}
M_{\ell \ell'}^-
	&= \frac{128 m_f^2}{\sqrt{(\ell^2-1)(\ell'^2-1)}}  \int_0^1 \frac{dx}{\pi} {C_{\ell-2}^2(1-2x)C_{\ell'-2}^2(1-2x)}\sqrt{x(1-x)}\\
	&= \frac{8}{3}m_f^2 \min(\ell,\ell') \sqrt{\frac{\min(\ell,\ell')^2 - 1}{\max(\ell,\ell')^2 -1}} 
	\delta_{\ell \ell' {\rm mod} 2}\, .
\end{split}
\end{equation}
%

%======================================================================%
%======================================================================%
%======================================================================%
\subsection{Spin even sector and $\langle TT\rangle$}\label{app_spin_even}

In Sec.~\ref{sec:MassDiagonalization} the matrix $\mathcal M = P^\mu P_\mu$ for the massive free fermion theory was diagonalized in the spin odd sector, which allowed us to compute current spectral densities in Sec.~\ref{sec:FreeFermionCorrelators}. In this section we obtain stress tensor spectral densities by similarly diagonalizing the spin odd sector. The stress tensor two-point function is constrained by Lorentz invariance and Ward identities to take the form \cite{Closset:2012vp} 
\begin{equation}\label{eq_TT}
\begin{split}
\langle T_{\mu\nu} T_{\alpha\beta}\rangle
	&= \frac{\tau_g}{|p|} P_{\mu\nu}P_{\alpha\beta} + 
	\frac{\tau_g'}{|p|} (P_{\mu\alpha}P_{\nu\beta} + (\mu \leftrightarrow \nu)) \\ 
	&\ + {\rm sgn}(m_f)\frac{\kappa_g}{192\pi} \left(\varepsilon_{\mu\alpha \lambda}p^\lambda P_{\nu\beta} + {\rm perm.}\right)\, ,
\end{split}
\end{equation}
where $P_{\mu\nu}\equiv p^2\eta_{\mu\nu} - {p_\mu p_\nu}$. Here $\tau_g$, $\tau_g'$ and $\kappa_g$ are functions of $p^2$ in general, except in a CFT where they are constants and $\tau_g=-\tau_g'$. These functions can be obtained in the Lagrangian approach by computing a fermion loop. This is done in Ref.~\cite{Bonora:2016ida}, which found%
	\footnote{Note that $T_{\mu\nu}^{\rm here} = 2 T_{\mu\nu}^{\rm there}.$}
\begin{subequations}\label{eq_taukappa_g}
\begin{align}
\tau_g
	&= \frac{i}{64\pi \gp^3} \left[4 + \gp^2 - \frac{(\gp^2 - 4)^2}{2\gp}{\rm atanh}\frac{\gp}{2}\right]\\
\tau_g'
	&= \frac{i}{64\pi \gp^3} \left[4 - \gp^2 + \frac{\gp^4 - 16}{2\gp}{\rm atanh}\frac{\gp}{2}\right]\\
\kappa_g
	&= 
	-\frac{3}{\gp} \left[\left(1-\frac{4}{\gp^2}\right){\rm atanh}\frac{\gp}{2} + \frac{2}{\gp}\right]\, .
\end{align}
\end{subequations}
In the kinematic regime $p_\perp=0$, we can extract these coefficients from 
\begin{subequations}
\begin{align}
\langle T_{--}T_{--}\rangle
	&= \frac{p_-^4}{|p|} \left(\tau_g + 2 \tau_g'\right) \, ,\\
\langle T_{-\perp}T_{-\perp}\rangle
	&= {p_-^2}{|p|}  \tau_g'\, ,\\
\langle T_{--}T_{-\perp}\rangle
	&= -{\rm sgn}(m_f){p_-^3}  \frac{\kappa_g}{96\pi}\, .
\end{align}
\end{subequations}
The stress tensor in the fermion theory is given by%
	\footnote{We consider the improved stress tensor (as opposed to the Noether current that generates translations) because only if it is symmetric will its correlator take the form \eqref{eq_TT}. } 
$T_{\mu\nu} = i\bar\Psi \gamma_{(\mu}\stackrel{\leftrightarrow}{\d}_{\nu)}\Psi$, the relevant components are
\begin{subequations}
\begin{align}
T_{--} &= \sqrt{2}i \left(\d_- \psi^* \psi - \psi^* \d_-\psi\right) \\
T_{-\perp}
	&= \frac{3i}{2\sqrt{2}} \d_\perp \psi^* \psi + \frac{i}{2\sqrt{2}}\d_-\psi^* \frac{\d_\perp-m_f}{\d_-} \psi
	+ \hbox{c.c.}\, .
\end{align}
\end{subequations}
The overlaps with the Dirichlet states are given by
\begin{subequations}
\begin{align}
\langle T_{--}(0) | \ell,+\rangle
	&=\frac{1}{8} \frac{\sqrt{p_-^{3}}}{\sqrt{\mu}}\delta_{\ell,2}\\
\langle T_{-\perp}(0) | \ell,-\rangle
	&= \frac{1}{4} \sqrt{\mu p_-} \left[\frac{\delta_{\ell 0 {\rm mod}2}}{\sqrt{\ell^2 - 1}} - \frac{\sqrt{3}}{2}\delta_{\ell 2}\right]\\
\langle T_{-\perp}(0) | \ell,+\rangle
	&=\frac{1}{2} \frac{\sqrt{p_-}}{\sqrt{\mu}}(-im_f)\delta_{\ell 0 {\rm mod}2} \, .
\end{align}
\end{subequations}
%

%======================================================================%
\subsubsection*{Diagonalization}
The parity-even, spin even sector spanned by the states $|\ell,+\rangle$, $\ell = 2,4,\ldots , \ell_{\rm max}$ is diagonalized $\mathcal M|\bar\alpha\rangle=\mathcal M_{\bar\alpha}|\bar\alpha\rangle$ by the states
\begin{equation}
|\bar\alpha\rangle
	= \sum_{j=1}^{\frac{\ell_{\rm max}}{2}}e_j^{(\bar\alpha)} |2j,+\rangle\, , 
	\qquad\hbox{with} \qquad
	e_j^{(\bar\alpha)} = \frac{2}{\sqrt{\ell_{\rm max}+1}}(-1)^j \sin \frac{2\pi \bar\alpha j}{\ell_{\rm max}+1}\, ,
\end{equation}
for $\bar\alpha = 1,2,\ldots,\ell_{\rm max}/2$. The corresponding eigenvalues are
\begin{equation}
\mathcal M_{\bar\alpha}
	= \mu^2 + 4 m_f^2 \sec^2 \phi_{\bar\alpha}\, ,
\end{equation}
with $\phi_{\bar\alpha}=\frac{\pi \bar\alpha}{\ell_{\rm max}+1}$. 

The parity-odd, spin even sector is spanned by the Dirichlet states $|\ell,-\rangle$, $\ell = 2,4,\ldots , \ell_{\rm max}$ and is diagonalized $\mathcal M|\bar\beta\rangle=\mathcal M_{\bar\beta}|\bar\beta\rangle$ by the states
\begin{equation}
|\bar\beta\rangle
	= \sum_{j=1}^{\frac{\ell_{\rm max}}{2}}e_j^{(\bar\beta)} |2j,+\rangle\, .
\end{equation}
for some $e_j^{\bar \beta}$; the exact form will not be needed.
The corresponding eigenvalues are
\begin{equation}
\mathcal M_{\bar\beta}
	= \mu^2 + 4 m_f^2 \sec^2 \phi_{\bar\beta}\, .
\end{equation}
The overlaps in the diagonal basis are given by
\begin{subequations}
\begin{align}
\langle T_{--}(0) | \bar\alpha\rangle
	&=-\frac{1}{4} \frac{\sqrt{p_-^{3}}}{\sqrt{\mu}}\frac{\sin2\phi_{\bar\alpha}}{\sqrt{\ell_{\rm max}+1}} \, ,\\
\langle T_{-\perp}(0) | \bar\beta\rangle
	&= - \frac{1}{4}\sqrt{\mu p_-} \frac{\cos 2 \phi_{\bar\beta}}{\sqrt{\ell_{\rm max}+1}}\, ,\\
\langle T_{-\perp}(0) | \bar\alpha\rangle
	&=\frac{im_f}{2} \frac{\sqrt{p_-}}{\sqrt{\mu}}\frac{\tan \phi_{\bar\alpha}}{\sqrt{\ell_{\rm max}+1}}\, .
\end{align}
\end{subequations}
%

%======================================================================%
\subsubsection*{Spectral densities}

We are now ready to compute the spectral densities using Eq.~\eqref{eq_spec_dens2}. From the overlaps above, these are found to be
\begin{subequations}
\begin{align}
\pi\rho_{T_{--}T_{--}}(p)
	&= \frac{1}{\ell_{\rm max}+1} \sum_{\bar\alpha} \frac{\Theta(\mu_{\bar\alpha}^2)}{\mu_{\bar\alpha}} \sin^2 2\phi_{\bar\alpha} \cdot \left(\frac{p_-^4}{16}\right)\, ,\\
\pi\rho_{T_{--}T_{-\perp}}(p)
	&= \frac{1}{\ell_{\rm max}+1} \sum_{\bar\alpha} \frac{\Theta(\mu_{\bar\alpha}^2)}{\mu_{\bar\alpha}} \sin 2\phi_{\bar\alpha} \tan \phi_{\bar\alpha} \cdot \left(\frac{im_fp_-^3}{8}\right) \, ,\\
\pi\rho_{T_{-\perp}T_{-\perp}}(p)|_{\rm even\, cut}
	&= \frac{1}{\ell_{\rm max}+1} \sum_{\bar\alpha} \frac{\Theta(\mu_{\bar\alpha}^2)}{\mu_{\bar\alpha}}\tan^2 \phi_{\bar\alpha} \cdot \left(\frac{m_f^2p_-^2}{4}\right) \, ,\\
\pi\rho_{T_{-\perp}T_{-\perp}}(p)|_{\rm odd\, cut}
	&= \frac{1}{\ell_{\rm max}+1} \sum_{\bar\beta} \mu_{\bar\beta} \Theta(\mu_{\bar\beta}^2)\cos^2 2\phi_{\bar\beta} \cdot \left(\frac{p_-^2}{16}\right)\, .
\end{align}
\end{subequations}
In the $\ell_{\rm max}\to \infty$ limit, one finds
\begin{subequations}
\begin{align}
\pi\rho_{T_{--}T_{--}}(p)
	&\to \frac{p_-^4}{16 |m_f|} \int_0^{\frac{1}{\pi} {\rm acos}\frac{2}{\gp}} dx\, \frac{\sin^2 2\pi x}{\sqrt{\gp^2 - 4 \sec^2 \pi x}}
	= \frac{p_-^4}{|p|}
	\frac{\gp^4 + 8 \gp^2 - 48}{64\gp^4} \, ,\\
\pi\rho_{T_{--}T_{-\perp}}(p)
	&\to -\frac{im_fp_-^3}{8|m_f|} \int_0^{\frac{1}{\pi} {\rm acos}\frac{2}{\gp}} dx\, \frac{\sin 2\pi x \tan \pi x}{\sqrt{\gp^2 - 4 \sec^2 \pi x}}
	= -i\frac{{\rm sgn}(m_f)p_-^3}{96\pi}6\pi\frac{\gp^2 - 4}{\gp^3} \, ,
\end{align}
\end{subequations}
Similarly adding the parity-even and odd cuts one finds
\begin{equation}
\pi \rho_{T_{-\perp}T_{-\perp}}
	\to  {p_-^2|p|} \frac{\gp^4 - 16}{64\gp^4}\, .
\end{equation}
The result from truncation is therefore
\begin{equation}
\Re\tau_g
	= -\frac{(\gp^2-4)^2}{64 \gp^4}\, , \qquad
\Re\tau_g'
	= \frac{\gp^4  -16}{64 \gp^4}\, , \qquad
\Im\kappa_g
	= 6\pi\frac{\gp^2 -4}{\gp^3}\, .
\end{equation}
These results exactly agree with the Lagrangian predictions \eqref{eq_taukappa_g}.

%######################################################################%
%======================================================================%
%======================================================================%
%======================================================================%
%######################################################################%
\section{Dirichlet basis from dimensional regularization}
\label{sec:DimRegDirichlet}

The Dirichlet basis was constructed in appendix \ref{app_dirichlet} by computing matrix elements with a sharp IR momentum cutoff $x = \frac{p_-}{P_-} \in [\epsilon, 1-\epsilon]$, and projecting to the kernel of the divergent part of this matrix. For free CFTs, this regularization scheme can be generalized to states of higher particle number by cutting off the momentum of each particle. However, this prescription does not generalize to generic CFTs, motivating the study of alternative regularization schemes that can be more readily generalized. In this section we study the theory using dimensional regularization (dim.~reg.) $dp_- \to d^{1+\epsilon'}p_- = p_-^{\epsilon'} dp_-$. We perform this replacement in the mode expansion
\begin{equation}
\psi(x) = \frac{1}{2^{1/4}} \int \frac{d^2 p}{(2\pi)^2} p_-^\epsilon e^{ipx} \psi_p\, .
\end{equation}
Two particle states now take the form (ignoring overall factors)
\begin{equation}
\begin{split}
|\phi,P\rangle
	&= \int d^3 y \, e^{-iyP} \left[f_\phi(-i\d^1,-i\d^2) :\psi^*(y_1)\psi(y_2):|_{y_{1,2}\to y}\right] |0 \rangle\\
	&= \int_0^1 \frac{dx}{\pi} [x(1-x)]^{\frac{1+\epsilon'}{2}}\sum_{p_\perp}f(p,P-p) :\psi^*_p \psi_{P-p}: |0\rangle.
\end{split}
\end{equation}
This replacement will simply add a factor of $[x(1-x)]^{\epsilon'}$ in any inner product or matrix elements. Let us now look at the matrix elements of the mass term in the parity-odd sector of primary states. These are given by \eqref{eq_Mminus} and are all divergent. A sharp momentum cutoff on the integral gave
\begin{equation}\label{eq_IRcutoff}
M_{\ell \ell'}^- = \frac{8  m_f^2 }{2^{(\delta_{\ell 0}+\delta_{\ell' 0})/2}} 
	\left[\frac{1}{\sqrt{\epsilon}} - \max(\ell,\ell')\right] \delta_{\ell \ell'\rm mod 2} 
	\qquad\qquad \hbox{(IR cutoff $\epsilon$)}\, .
\end{equation}
If instead the integral in \eqref{eq_Mminus}  is evaluated with the dim.~reg.~prescription explained above, assume $\epsilon'>1/2$ and then analytically continue to $\epsilon' \to 0$ one finds
\begin{equation}\label{eq_dimreg}
\quad
M_{\ell \ell'}^- = \frac{8  m_f^2 }{2^{(\delta_{\ell 0}+\delta_{\ell' 0})/2}} 
	\left[0 - \max(\ell,\ell')\vphantom{\frac12}\right] \delta_{\ell \ell'\rm mod 2} 
	\qquad\quad \hbox{($d^{1+\epsilon'}p_-$ dim.~reg.)}\, .
\end{equation}
The effect of dim.~reg.~is simply to remove the power law divergence in \eqref{eq_IRcutoff}. In the limit $\ell_{\rm max}\to \infty$ the spectra in both schemes agree, except that the infinite eigenvalues $\sim \frac{1}{\sqrt{\epsilon}}$ in \eqref{eq_IRcutoff} are replaced by negative eigenvalues in \eqref{eq_dimreg}. Removing the corresponding eigenvectors, one recovers the Dirichlet basis with the dim.~reg.~prescription. At finite $\ell_{\rm max}$ the spectra and corresponding bases will however differ in both schemes.

%######################################################################%
%======================================================================%
%======================================================================%
%======================================================================%
%######################################################################%
\section{Connection to Covariant Formulation}
\label{sec:bs}

In this section we show how to relate the Bethe-Salpeter equation, which 
is used to compute the correlators, to the Hamiltonian equation. To warm up,
we will consider the $O(N)$ model first, and express the Hamiltonian equation
in the conformal basis. Next, we will show how to derive the Hamiltonian equation
from the Bethe-Salpeter equation in the large $N_f$ Chern-Simons theory.

\subsection{Large $N$ $O(N)$}

We will begin as a warm-up with the $O(N)$ model at large $N$.  The Bethe-Salpeter equation in this case is
\be
\psi(p,p',r) &=& S(p) S(p-r) \delta(p-p') +\lambda  \int d^d k \psi(p,p',r) S(p) S(p-r)
\ee
We now follow steps similar to 't Hooft \cite{tHooft:1974pnl}. 
However, instead of considering directly the homogenous equation for the 
``blob'' as in 't Hooft, we will not integrate entirely over $d ^3 p'$, so that 
we have an equation for the full correlator.\footnote{A physical motivation to 
consider the full correlator is that the states $\phi_i$ in the $O(N)$ model 
are physical states, unlike the quarks in the $2D$ 't Hooft model which are 
confined.}
Next, we consider
\be
\phi(\vec{p}, \vec{p}',r) &=& \int dp_+ dp'_+ \psi(p,p',r)
\ee
Which satisfies the equation
\be
\label{eq:bson}
\phi(p,p',r) &=& \left( \int dp_+ S(p)S(p-r) \right) \left( \delta^2(p-p') + \lambda \int d^2 k \phi(p',k,r) \right)
\ee
This correlator can be used to compute the two-point functions between the bilinear 
currents, which do not mix with higher currents at large N. 
We can indentify the currents symbolically in terms of a discrete set of parameters,
\be
\braket{J_{\alpha}(q) J_{\alpha}(-q)}
&=& \int d^2 p d^2 p' F_{\alpha}^{(2)}(p) \phi(\vec{p}, \vec{p}', q) 
F_{\alpha'}^{(2)}(p')
\,.
\ee
where the $F$'s are the wavefunctions of the bilinear states, 
$F_{\alpha}^{(2)}(p) \equiv \braket{p | J_{\alpha}}$.\footnote{Explicitly,
the wavefunctions are proportional to Gegenbauer polynomials,
\begin{align}
F_{\ell -}(p) &= C_\ell \left(1 - 2 \frac{p_-}{P_-} \right) 
\\ F_{\ell \perp}(p) &= \frac{p_\perp}{P_-} C_{\ell-1}^1
\left(1 - 2 \frac{p_-}{P_-}\right) \,.
\end{align}
}
In this model, the parameters $\alpha$ just denote the spin and parity 
quantum numbers.

We will now show how the Bethe-Salpeter equation \eqref{eq:bson} can be related 
to the conformal truncation Hamiltonian equation, and how the spectral density
can be computed in terms of the wavefunctions of the Hamiltonian.
To show that, we first perform the $d p_+$ integral in the first term in 
\eqref{eq:bson} explicitly,\footnote{The scalar propagators each decay like $p_+^{-1}$, so $S(p) S(p-r)$ decays like $p_+^{-2}$ and the $dp_+$ integral can be done by residues:
\be
\int dp_+ S(p)S(p-r) &=& \int dp_+ \frac{1}{(2p_+ p_- - p_\perp^2 -m^2 + i\epsilon)(2(p_+ - r_+)(p_- - r_-) - p_\perp^2 -m^2+i \epsilon)} \nn\\
 &=& \theta(p_-) \theta(r_- - p_-) \frac{ \pi i r_- }{p_- (r_--p_-)} \left( r^2 - \frac{p_\perp^2 +m^2}{p_-(r_--p_-)}r_-^2 \right)^{-1} \nn\\
  &=& \theta(x) \theta( 1-x) \frac{ \pi i  }{x (1-x)} \left( r^2 - \frac{p_\perp^2 +m^2}{x(1-x)} \right)^{-1} 
 \ee
In the last line, we set $r_-=1$ and $p_- = x$. } to obtain
\be
 \left( r^2 - \frac{p_\perp^2 +m^2}{x(1-x)} \right) \phi(p,p',r) &=& \theta(x) \theta( 1-x) \frac{ \pi i  }{x (1-x)}  \left( \delta^2(p-p') + \lambda \int d^2 k \phi(k,p',r) \right) \nn\\
\ee
where we have taken $r_\perp=0, r_-=1$, and defined $x=p_-/r_-$. 
We define the action of the non-perturbative squared mass operator as
\be
M^2[f(\cdot)](p) &\equiv& \frac{p_\perp^2+m^2}{x(1-x)} f(x) 
+ \lambda \theta(x) \theta( 1-x) \frac{ \pi i  }{x (1-x)} \int d^2 k f(k)
\ee
we therefore have
\be
\label{eq:IntONBE}
M^2[\phi(\cdot,\vec{p}',r) ](p)  &=& -\frac{i \pi}{x(1-x)} \delta^2(p-p')
+r^2 \phi(p,\vec{p}',r)
\ee
Let $\phi_{E,n}(\vec{p})$ be the solutions of the homogeneous part of this equation:
\be
\label{eq:hamonhom}
M^2[\phi_{E,n}](p) = E^2 \phi_{E,n}(\vec{p})
\ee
where $E$ is a continuous energy parameter, while $n$ denotes a countable 
set of quantum numbers.
These solutions will the orthogonal, and we normalize them so that
\be
\sum_n\int_{E_n} d E^2 \phi_{E,n}(\vec{p}) \phi_{E,n}^*(\vec{p}') = \frac{1}{x(1-x)} \delta^2(p-p')
\ee
where $E_n$ is the lower bound on the energy of states with index $n$.
Then, we can construct $\phi(\vec{p},\vec{p}',q) $ as
\be
\phi(\vec{p},\vec{p}',q) &=& i \pi \sum_n\int_{E_n} \frac{dE^2}{q^2-E^2+i \epsilon} \phi_{E,n}(\vec{p}) \phi_{E,n}^*(\vec{p}')
\ee
It is straightforward to see that this construction satisfies (\ref{eq:IntONBE}). 
Furthermore, the spectral density is just the real part of these correlators
\be
\label{eq:onspecdens1}
\rho_{\alpha \alpha'} &=& {\rm Re}\braket{J_\alpha(-q) J_{\alpha'}(q)} = 
\int d^2 p d^2p' F_\alpha(p) F_{\alpha'}(p') {\rm Re}(\phi(\vec{p},\vec{p}',q)) \nn\\
 &=& \pi^2 \sum_{n | E_n^2 < q^2} 
\left( \int d^2 p  \phi_{|q|,n} (\vec{p}) F_\alpha(p) \right)
\left( \int d^2 p'  \phi_{|q|,n} (\vec{p}') F_{\alpha'}(p') \right)^*
\ee

Finally, we want to show explicitly that the Hamiltonian equation \eqref{eq:hamonhom} 
is equivalent to the Hamiltonian equation in the conformal basis. 
To do so, we define the states $\ket{E,n}$ via 
$\phi_{E,n}(p) \equiv \frac{1}{p(1-p)} \braket{p | E,n}$
Next, we project Eq. \eqref{eq:hamonhom} onto the conformal basis 
$\ket{\mu, \ell}$, by integrating it with the kernel  
$\int d^2 p \braket{p | \mu,\ell}$, where 
\begin{equation}
\braket{p| \ell,\mu} =  
f_\ell(x) \delta\left(\mu^2 - \frac{p_\perp^2}{2 x (1-x)}\right)
\end{equation}
is the representation of the conformal basis states in the parton momentum space,
$f_\ell(x)$ being a orthonormal set of functions in the interval $[0,1]$ with measure
$\frac{1}{\sqrt{x(1-x)}}$. Here we only parity-invariant states, but the extension
to parity-odd states is straightforward.
Taking into account the completeness relation 
$\sum_{\ell} \int d \mu^2 \mu \ket{\mu,\ell} 
\bra{\mu,\ell} = \mathbb{I}$, we arrive at
\begin{align}
\label{eq:bsonconf}
E^2 \phi_{E,n}(\mu, \ell) &= \mu^2 \phi_{E,n}(\mu, \ell) 
+m^2 \sum_{\ell'} \int d x x^{-\frac{3}{2}} (1-x)^{-\frac{3}{2}} f_\ell(x)
f_{\ell'}(x) 
\\ &+ \lambda \left(\int d x \frac{1}{\sqrt{x(1-x)}} f_{\ell}(x) \right) 
\int d \mu'^2 \sum_{\ell'} \frac{1}{\sqrt{\mu \mu'}} 
\left(\int d x \frac{1}{\sqrt{x(1-x)}} f_{\ell'}(x) \right)
\phi_{E,n}(\mu', \ell')
\end{align}
where we defined $\phi_{E, n}(\mu, \ell) \equiv \sqrt{\mu} 
\braket{\mu,\ell | E,n}$.
Equation \eqref{eq:bsonconf} is equivalent to the conformal Hamiltonian 
truncation equation, leading to the matrix elements in section \ref{sec:onmodel}.
Finally, the spectral densities \eqref{eq:onspecdens1} can be expressed as a sum over these 
wavefunctions,
\begin{align}
\rho_{\alpha \alpha'}(q^2) &= \sum_{n | E_n < q^2} 
\left(\sum_{\ell} \int d x \frac{1}{\sqrt{x(1-x)}} F_{\alpha}(x) f_{\ell}(x) 
\int d \mu^2 \phi_{|q|, n}(\mu,\ell)\right)
\\ & \left(\sum_{\ell'} \int d x' \frac{1}{\sqrt{x'(1-x')}} F_{\alpha'}(x') f_{\ell'}(x') 
\int d \mu^2 \phi_{|q|, n}(\mu,\ell)\right)^*
\end{align}
where we specialized to the parity-even sector, but the generalization 
is straightforward.
In this model, the spectral density is proportional to $\delta_{\ell \ell'} 
\delta_{\alpha \alpha'}$, however the procedure to extract the correlators 
and spectral densities from $\phi$ is general.

\subsection{Large $N_f$ CS}

\begin{figure}[t!]
\label{fig:bs}
\begin{center}
\includegraphics[width=0.6\textwidth]{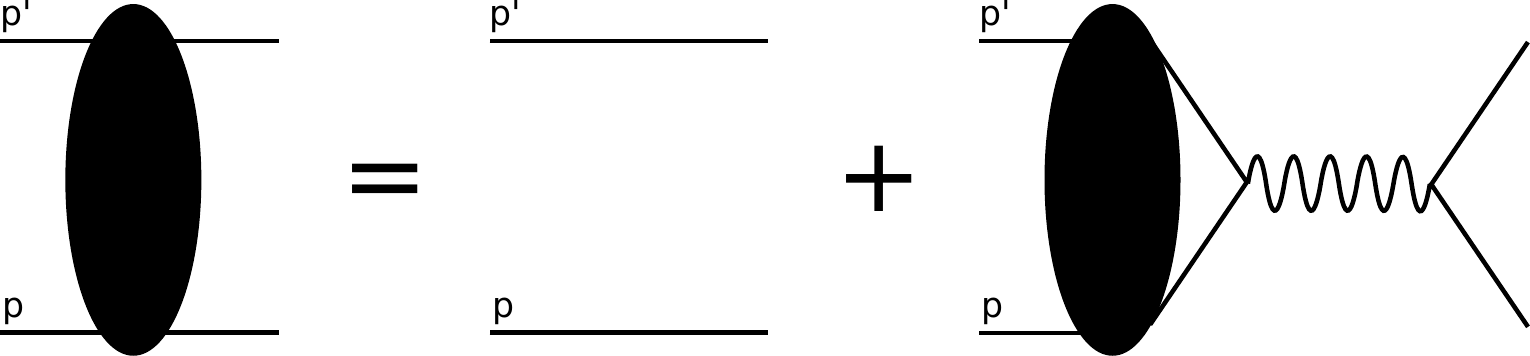}
\caption{Symbolic Bethe-Salpeter equation for the large $N_f$ CS theory}
\end{center}
\end{figure}

In a similar way, we can show that the Bethe-Salpeter equation for the 
large $N_f$ Chern-Simons theory can be cast as the Hamiltonian equation for the 
wavefunctions. There is, however, a subtlety related to the presence of 
unphysical degrees of freedom on the light-cone.
In LC gauge, our Lagrangian is
\be
\CL = \bar{\Psi} (i \slashed{\partial} + m_f) \Psi - a_i \bar{\Psi} \gamma^i \Psi + \frac{\kappa}{\pi} a_+ \partial_- a_\perp - m_a a_\perp^2
\ee
We will write the tree-level gauge boson propagator as
\be
\< a_\mu (p) a_\nu(-p) \> = D_0(p) &=& \frac{i}{p_-} \left( \begin{array}{cc} -\frac{\lambda^2 m_a}{p_-} & \lambda \\ -\lambda & 0 \end{array} \right)
\ee
where $\lambda = \pi/\kappa$. Similarly to the previous subsection, 
we consider the Bethe-Salpeter equation for the 
``transfer matrix'' $\psi_{b c}^{a d} (p, p',r)$ (see figure \ref{fig:bs})
or equivalently the resummed correlator in momentum space 
$\<\bar{\Psi}_i(p') \Psi^i(p'-r) \bar{\Psi}_j(p) \Psi^j(p-r)\>$, with the overall
momentum delta function factored out.
Let $S(p)$ be the fermion propagator.  Then, the Bethe-Salpeter equation can be written
as
\be
\label{eq:bs1}
\psi_{b c}^{a d}(p,p',r) &=& S^a_c(p) S_b^d(p-r)\delta^3(p-p') 
- \int d^3 k \ \tr\left[ \psi^{a e}_{b f}(k,p',r) 
(\gamma^\nu)_e^f \right] D_{0,\nu \rho}(r) (S(p) \gamma^\rho S (p-r))_c^d \nn\\
\ee
We define the integrated transfer matrix $\phi_{b c}^{a d}$ as
\be
\phi_{b c}^{a d} \equiv \int dp_+ dp_+'  \psi_{b c}^{a d}(p,p',r)
\ee
To derive the Bethe Salpeter equation for $\phi$ we need to evaluate the tensor
\be
(\phi_0)_{b c}^{a d}(p,r) \equiv \int dp_+ S^a_c(p) S_b^d(p-r)
\ee
To simplify the analysis, we will split the propagator into a ``regular term'', which is analytic in $p^+$, and 
an ``on-shell'' term, where $p_+$ is evaluate at the single particle pole,
\begin{align}
S(p) &= \frac{p_\mu \gamma^\mu + m_f \mathbb{I}}{p^2 - m_f^2 + i\epsilon} = S_{\rm OS}(p) + \frac{1}{2 p_-} \gamma^+
\\ S_{\rm OS}(p) &\equiv \frac{1}{\sqrt{2}} \frac{u(p) \bar{u}(p)}{p^2 - m_f^2 + i \epsilon} 
=  \frac{\frac{m_f^2+p_\perp^2}{2 p_-}\gamma^+ + p_- \gamma^- + p_\perp \gamma^\perp
+ m_f \mathbb{I}}{p^2 - m_f^2}
\end{align}
 where the fermion wavefunction $u(\vec{p})$ is the usual solution in 3d, with $p_+ = (p_+)_{\rm on-shell}$: 
\be
u(\vec{p}) &=& \sqrt{p_- } \left(\begin{array}{c} \frac{ip_\perp-m}{\sqrt{2} p_-} \\ -1 \end{array} \right)
\ee
and $\bar{u} = u^* \left(\begin{array}{cc} 0 & 1 \\ 1 & 0 \end{array} \right)$,
so that $(\slashed{p}+m) u(\vec{p}) = 2 m_f u(\vec{p})$.
$S_{\rm OS}$ is proportional to a projection operator on a one-dimensional subspace 
of the space of spinors. Consequently, we have two contributions to $\phi$,
\be
(\phi_0)_{b c}^{a d}(p,r) \equiv 
(\phi_{\rm OS})_{b c}^{a d}(p,r) + (\phi_{\rm reg})_{b c}^{a d}(p,r)
\ee
The right prescription, from comparison of the final correlators with a Feynman diagram calculation, turns out to be to discard the ``regular term'' in the derivation of 
the Bethe-Salpeter equation. This prescription intuitively corresponds to discarding 
the unphysical degrees of freedom on the light cone, though we do not have an a priori derivation for it.
Performing the integral we obtain
\begin{align}
& (\phi_{\rm OS})_{b c}^{a d}(p,r) = \frac{\sqrt{2}}{r_-} 
\frac{i \pi}{x(1-x)} \theta(x)\theta(1-x) \frac{1}{E_0^2}  u_b(p-r) u_c(p) \bar{u}^a(p) \bar{u}^d(p-r)
\\ & x \equiv \frac{p_-}{r_-} \,, \qquad E_0 \equiv 2 r_+ r_- - \frac{p_\perp^2 + m^2}{x(1-x)}
\end{align}
Since the external momentum $r^\mu$ just sets the momentum frame we are working with, we will set $r_\perp = 0, r_-=1, r_+ = \frac{1}{2} \mu^2$, and for conciseness we will not write the $\mu$-dependence explicitly.
Integrating the equation \eqref{eq:bs1}, we get
\begin{align}
\label{eq:bs2}
& \phi_{b c}^{a d}(p,p',r) = \frac{1}{E_0^2(p)} 
\sqrt{2} \frac{i \pi}{x(1-x)} \theta(x) \theta(1-x)
u_c(p) \bar{u}^d(p-r) 
\\ & \left[u_b(p'-r) \bar{u}^a(p')  \delta^2 (p-p')
- \int d^2 k \ \left[ \phi^{a e}_{b f}(k,p',r) 
(\gamma^\nu)_e^f \right] D_{0,\nu \rho}(r) (\bar{u}(p) \gamma^\rho u(p-r)) \right]
\end{align}
Taking the contraction of $\phi$ with $\gamma$ matrices, and integrating in $\int d^2 p d^2 p'$, will just give 
the expression for the correlators. However, we will not do this, since we seek an equation for the wavefunctions.
Formally, the solution can be found by first solving the homogeneous equation,
\begin{equation}
M^2[\phi_{E,n}]_c^d = E^2 (\phi_{E,n})_c^d
\end{equation}
where we defined the squared mass operator
\begin{align}
& M^2[f(.)]_c^d(p) \equiv E^2_0(p) f_{c}^{d}(p) 
\\ &+ \sqrt{2} \frac{\pi i}{x(1-x)} \theta(x)\theta(1-x)
u_c(p) \bar{u}^d (p-r)
\int d^2 k f_{e}^{f}(k) (\gamma^\nu)_f^e D_{0,\nu \rho}(r)
(\bar{u}(p)\gamma^\rho u(p-r)) \,.
\end{align}
The wavefunctions will be normalized so that
\begin{equation}
\sum_\alpha \int d E^2 (\phi_{E,\alpha})^d_c(p) (\phi_{E,\alpha}^*)_b^a(p') =
\sqrt{2} \frac{i \pi}{x(1-x)}
u_c(p) \bar{u}^d(p-r) u_b(p'-r) \bar{u}^a(p') \delta^2(p-p')
\end{equation}
Therefore, the solution of \eqref{eq:bs2} can be written as
\begin{equation}
\phi^{a d}_{b c}(p,p',r) = i \pi \sum_{n} \int_{E_n} d E^2 
\frac{1}{r^2 - E^2 + i \epsilon}
(\phi_{E,n})^d_c(p) (\phi_{E,n}^*)_b^a(p')
\end{equation}
from which the spectral densities for the currents $J_-$, $J_\perp$ can be computed as
\begin{equation}
\rho_{\mu\nu}(q) = {\rm{Re}} \int d^2 p d^2 p'
(\gamma^\mu)^b_a \phi^{a d}_{b c}(p,p',r) (\gamma^\nu)^c_d \,.
\end{equation}

%######################################################################%
%======================================================================%
%======================================================================%
%======================================================================%
%######################################################################%
\section{Large $N_f$ CS at $m_f=0$ in Primary State Basis}
\label{app:NonDirichlet}

In this appendix we study the massless theory $m_f= 0$, which can be addressed using the primary basis  \eqref{eq:masslessbasis} instead of the Dirichlet basis. The matrix elements of the interaction are given in \eqref{eq_M4_massless_el}. Since the interaction only acts on the states $|\ell,s,\mu\rangle = |1,\pm,\mu\rangle$, the Hamiltonian is already diagonal except in the subspace spanned by these two states where it is given by (we drop the spin index since $\ell=1$ in this subspace)
\begin{equation}
\langle \pm,\mu' | \mathcal M^{\rm int} | \pm,\mu \rangle
	= \left(\begin{array}{cc}
	-\frac{\pi^2\lambda^2 m_a}{2\sqrt{\mu\mu'}}&i\frac{\pi\lambda}{8}\sqrt{\mu/\mu'}\\
	-i\frac{\pi\lambda}{8}\sqrt{\mu'/\mu}&0\\
	\end{array}\right) \, ,
\end{equation}
Using the overlaps of the current operators with the basis states
\begin{equation}
\langle j_-|+,\mu\rangle = \frac{1}{4\sqrt{\mu}}\, ,\qquad
\langle j_\perp|-,\mu\rangle = \frac{\sqrt{\mu}}{4}\, ,
\end{equation}
the full mass matrix in this subspace can be written
\begin{equation}\label{eq_M_op_massless}
\mathcal M
	= \mu^2 \mathbbm 1 + 2\pi i \lambda \left(\vphantom{I^I}|j_-\rangle\langle j_\perp| - |j_\perp\rangle\langle j_-|\right) - 8m_a \pi^2\lambda^2 |j_-\rangle\langle j_-|\, .
\end{equation}
The rest of the diagonalization will closely follow Sec.~\ref{sec:BraKetNotation} where the $O(N)$ model was diagonalized, using Dirac notation. For simplicity in this section we will use dimensional regularization to perform the $\mu^2$ integrals, and we can therefore take $m_a=0$ (see Section \ref{sec_CS_Nf} and in particular Eq.~\eqref{eq_jpjp_div} for regularization using a sharp cutoff, where a counterterm $m_a\neq 0$ is needed). Looking for an eigenvector of \eqref{eq_M_op_massless} with eigenvalue $q^2$ leads to 
\begin{equation}
(q^2 - \mu^2)|\psi\rangle
	= 2\pi i\lambda \left(\vphantom{I^I}|j_-\rangle\langle j_\perp| - |j_\perp\rangle\langle j_-|\right)|\psi\rangle\, .
\end{equation}
This equation can be inverted up to a term in the kernel of $(q^2 - \mu^2)$ which can be written $|\psi\rangle^{(0)} = \sum_i C_i \hat D |j_i\rangle$, where $i=-,\perp$. One then finds
\begin{equation}
|\psi\rangle 
	= \sum_i C_i \hat D |j_i\rangle + S_i \tilde D |j_i\rangle\, ,
\end{equation}
where the coefficients $S_i$ satisfy
\begin{equation}
S_i
	= \sum_j V_{ij} \langle j_j | \hat D | j_j\rangle C_j\, ,
\end{equation}
where the non-zero entries of  $V$ are $V_{\perp-}^* = V_{-\perp} = 2\pi i\lambda$ (note that here the matrix $\langle j_i | \hat D | j_j\rangle\propto \delta_{ij}$ is diagonal and $\langle j_j | \hat D | j_j\rangle = 0$, unlike in the general case with $m_f\neq 0$ studied in section \ref{sec_CS_Nf}). The norm of the state is
\begin{equation}
\langle \psi|\psi\rangle
	= \sum_i \left(|C_i|^2 + \frac{1}{4}|S_i|^2\right) \langle j_i | \hat D | j_i \rangle\, .
\end{equation}
Let us focus on the solution with $C_\perp = 0$ and $C_-=1$. It has the form
\begin{equation}
|\psi,-\rangle
	= \hat D |j_-\rangle  - \tilde D |j_\perp \rangle2\pi i\lambda \langle j_- | \hat D| j_-\rangle\, ,
\end{equation}
with norm
\begin{equation}
\mathcal N_\psi \equiv 
\langle \psi|\psi\rangle 
	= \langle j_- | \hat D| j_-\rangle + (\pi \lambda)^2 \langle j_- | \hat D| j_-\rangle^2 \langle j_\perp | \hat D| j_\perp\rangle\, .
\end{equation}
Since the current $j_-$ only overlaps with this state, we can directly compute its spectral density
\begin{equation}
\pi \rho_{--}
	= \frac{|\langle j_- | \psi,-\rangle|^2}{\mathcal N_\psi}
	= \frac{1}{\langle j_- | \hat D| j_-\rangle^{-1}+ (\pi\lambda)^2 \langle j_\perp | \hat D| j_\perp\rangle}
	= \frac{1}{16 \mu} \frac{1}{1+ \left(\frac{\pi\lambda}{16}\right)^2}\, , 
\end{equation}
where in the last step we used $\langle j_- | \hat D| j_-\rangle = \frac{1}{16\mu}$ and $\langle j_\perp | \hat D| j_\perp\rangle = \frac{\mu}{16}$. This reproduces the known result (\ref{eq:FinalJJ}) at $m_f=0$.

%######################################################################%
%======================================================================%
%======================================================================%
%======================================================================%
%######################################################################%
\section{Large $N_f$ Interaction Matrix Elements}
\label{app:InteractingCSMatrixElements}

The general form of matrix elements of $\mathcal M_4$ in two-particle states was given in \eqref{eq_M4_matrixel}. This leads to the following matrix elements in the Dirichlet basis $\mathcal M^{\rm int}_{\ell' s',\ell s}\equiv \langle\ell',s' | \mathcal M_4 |\ell,s\rangle$: 
\begin{equation}
\begin{split}
\mathcal M^{\rm int}_{\ell' s',\ell s}
	&= - \delta_{s+}\delta_{s'+}\delta_{\ell 1}\delta_{\ell' 1} \frac{\pi^2 \lambda^2 m_a}{4\sqrt{\mu\mu'}}
	- \delta_{s+}\delta_{s'+}\delta_{\ell 1}\delta_{\ell' 1{\rm mod} 2}\frac{m_f}{\sqrt{\mu\mu'}} \frac{\pi \lambda}{2} \\
	&+ \delta_{\ell'1} \delta_{s'+} \delta_{s-} \delta_{\ell 1{\rm mod} 2}\frac{1}{\sqrt{\ell^2-1}} \frac{i\pi\lambda}{4} \sqrt{\frac{\mu}{\mu'}}  + \hbox{h.c.}\, .
\end{split}
\end{equation}
It will be convenient to work in the basis constructed in section \ref{sec:MassDiagonalization} that diagonalizes the mass term. In this basis, the matrix elements read 
\begin{subequations}\label{eq_M4_elements}
\begin{align}
\mathcal M^{m_a}_{\alpha\alpha'} 
	&= -\frac{2m_a\pi^2\lambda^2}{\sqrt{\mu\mu'}} \frac{1}{\ell_{\rm max}+1}\cos\phi_{\alpha+}\cos\phi_{\alpha'+}\, , \\
\mathcal M^{m_f}_{\alpha\alpha'}
	&= -\frac{m_f \pi\lambda}{\sqrt{\mu\mu'}} \frac{1}{\ell_{\rm max}+1} 
	 \frac{\cos\phi_{\alpha+}} {\cos\phi_{\alpha'+}} + (\alpha \leftrightarrow  \alpha')\, , \\
\mathcal M^{\rm CS}_{\alpha\beta}
	&=  \frac{i \pi\lambda}{2} \sqrt{\frac{\mu_\beta}{\mu_\alpha}}\frac{1}{\ell_{\rm max}+1} \cos\phi_{\alpha+} \sin\phi_{\beta-}\, .
\end{align}
\end{subequations}
We will particularly be interested in the spectral densities of the current $j_-,\,j_\perp$. In the free theory, their overlaps with the mass eigenstates $|\alpha\rangle$, $|\beta\rangle$ were given in \eqref{eq_j_alpha} and \eqref{eq_j_beta}. In the interacting theory, the overlaps are given by
\begin{equation}
\< \alpha , \mu |j_-\> =  c_\alpha(\mu) , \quad \< \beta, \mu | j_-\> =0, \quad 
\< \alpha , \mu |j_\perp\> = -2 i m_f a_\alpha(\mu), \quad \< \beta, \mu | j_\perp\> =  s_\beta(\mu) ,
\end{equation}
where we have defined the following vectors:
\be
 c_\alpha(\mu)  \equiv -\frac12 \mu^{-\frac{1}{2}} \frac{\cos \phi_{\alpha+} }{\sqrt{\ell_{\rm max}+1}}, \quad s_\beta(\mu) \equiv -\frac12 \mu^{\frac{1}{2}}  \frac{\sin \phi_{\beta-} }{\sqrt{\ell_{\rm max}+1}}, \quad a_\alpha(\mu) \equiv -\frac12 \mu^{-\frac{1}{2}} \frac{\sec \phi_{\alpha+}}{\sqrt{\ell_{\rm max}+1}}. \nn
\ee
Using this notation, the interaction with matrix elements \eqref{eq_M4_elements} is simply given by
\begin{equation}\label{eq_Mint_operator}
\mathcal M^{\rm int}
	= 2\pi i \lambda |j_-\rangle\langle j_\perp| + {\rm h.c.}
	- 8\pi^2\lambda^2 m_a |j_-\rangle\langle j_-|\, .
\end{equation}

\medskip

The matrices $\tilde{D}$ and $\hat{D}$ denote matrices whose diagonal components are the  principal value pole or $\delta$ function, respectively. 
For instance,
\be
\< \alpha, \mu | \tilde{D} | j_-\> = P.V. \frac{c_\alpha(\mu)}{\mu^2(q,\alpha)-\mu^2},
\ee
and
\be
\< j_- | \tilde{D} | j_-\> = \sum_\alpha \int\frac{d\mu^2}{2\pi}P.V. \frac{c_\alpha^2(\mu)}{\mu^2(q,\alpha)-\mu^2}.
\ee

We will also use the following matrix elements: 
\be
\< j_- | \tilde{D} | j_-\> &=& - \frac{1}{8(\ell_{\rm max}+1)} \sum_\alpha \llb -\mu^2(\alpha,q)\rrb^{-\frac{1}{2}} \cos^2 \phi_{\alpha +} , \\
\< j_\perp | \tilde{D} | j_- \> = \< j_- | \tilde{D} | j_\perp\>^* &=& 2 i m_f \frac{1}{8(\ell_{\rm max}+1)} \sum_\alpha \llb -\mu^2(\alpha, q) \rrb^{-\frac{1}{2}}, \nn\\
\< j_\perp | \tilde{D} | j_\perp\> &=& \frac{ 1}{8(\ell_{\rm max}+1)} \left( -\frac{\Lambda (\ell_{\rm max}-1)}{2\pi}  \right. \nn\\
 && \left. + \sum_\beta  \llb -\mu^2(\beta,q) \rrb^{\frac{1}{2}} \sin^2 \phi_{\beta-}  -4 m_f^2 \sum_\alpha \llb -\mu^2(\alpha,q) \rrb^{-\frac{1}{2}} \sec^2 \phi_{\alpha +}  \right) . \nn
\ee
The corresponding matrix elements for $\hat{D}$ were obtained in section \ref{sec:FreeTheory}.\footnote{ See equation (\ref{eq_spec_dens_sim2}) for the free theory spectral function $\rho^{\rm free}$, and the connection between $\hat{D}$ and $\rho^{\rm free}$ is (\ref{eq:DhatVsRho}).} In the limit $\ell_{\rm max} \rightarrow \infty$, the above expressions simplify to 
\be
\label{eq:jjtildeIntegrals}
 \< j_- | \tilde{D} | j_-\> &=&-\frac{1}{8m_f} \int_{\pi^{-1} \cos^{-1}(\frac{2}{\gp})}^{\frac{1}{2}} \frac{dx \cos^2 \pi x }{( 4  \sec^2\pi x-\gp^2)^{1/2}} = \frac{1}{8\pi m_f} \left( \frac{1}{\gp^2}-\frac{\left(\gp^2+4\right) \coth ^{-1}\left(\frac{\gp}{2}\right)}{2 \gp^3}\right) , \nn\\
\< j_\perp | \tilde{D} | j_-\> &=& \frac{i}{4}   \int_{\pi^{-1} \cos^{-1}(\frac{2}{\gp})}^{\frac{1}{2}} \frac{dx  }{( 4  \sec^2\pi x-\gp^2)^{1/2}} =  \frac{ i \coth ^{-1}\left(\frac{\gp}{2}\right)}{4\pi \gp}, \nn\\
\< j_\perp | \tilde{D}| j_\perp\> &=& - \frac{ \Lambda}{16\pi} - m_f  \frac{1}{8} \int_{\pi^{-1} \cos^{-1}(\frac{2}{\gp})}^{\frac{1}{2}} \frac{dx (4+ \gp^2 \sin^2 \pi x)  }{( 4  \sec^2\pi x-\gp^2)^{1/2}} -\frac{1}{2\pi} m_f  \log 2 \nn\\
 &=&\frac{1}{8\pi} \left[\left( - \frac{ \Lambda}{2} -(2+4\log 2)m_f\right)+ m_f \left(1 - \frac{\left(\gp^2+4\right) \coth ^{-1}\left(\frac{\gp}{2}\right)}{2   \gp}\right) \right]  .
 \ee
at $\gp>2$.\footnote{The $ \log 2$ term in the second-to-last line is a bit subtle because it is due to the difference between the $\ell_{\rm max} \rightarrow \infty$ limit of the sum, and the approximation of the sum as an integral over the summand. Shamefully, we discovered it numerically. It is easiest to derive analytically by evaluating the sum at $q=0$.  }  At $0< \gp <2$, the expressions are the same except with every $\coth^{-1}$ replaced by a $\tanh^{-1}$.  Recall that we define $\gp = \frac{q}{m_f}$.

%######################################################################%
%======================================================================%
%======================================================================%
%======================================================================%
%######################################################################%
\section{Wavefunction Normalizations}
\label{app:WvNm}

In our analysis, we often encounter eigenstates with wavefunctions of the form
\be
\psi_\alpha(\mu; q) &=& C_\alpha \delta(q^2 - \mu^2 -m_\alpha^2) + f_\alpha(\mu) P.V. \frac{1}{q^2-\mu^2 - m_\alpha^2},
\ee
where $f_\alpha$ is regular away from $\mu=0$.  These states are eigenstates of a Hamiltonian with a continuous set of eigenvalues, and therefore their norm receives only $\delta$ function contributions.  These contributions can come from the overlap of two $\delta$ function terms, as well as from the overlap of two principal value terms:
\be
\int d \mu^2 \psi_\alpha^*(\mu;q) \psi_\alpha(\mu,q') &=& \sum_\alpha  \Theta(q^2 - m_\alpha^2) \left[ C_\alpha C'^*_\alpha + \pi^2 f_\alpha(\sqrt{q^2-m_\alpha^2}) f'^*_\alpha(\sqrt{q^2 - m_\alpha^2}) \right] \nn\\ && \times \delta(q^2-q'^2).
\ee
The contribution from the principal value parts can be derived as follows.  First, recall the following representation of the principal value part:
\be
\lim_{\epsilon\rightarrow 0^+} \frac{ x}{x^2+\epsilon^2} = P.V. \frac{1}{x}
\ee
The normalization of $\psi_\alpha$ above involves two kinds of integrals over principal values:
\be
\int dx g(x) P.V. \frac{1}{x-a} \textrm{ and } \int dx g(x) P.V. \frac{1}{x-a} P.V. \frac{1}{x-b},
\ee
where $g(x)$ is regular.  We are interested only in divergences that arise in the integral from the region near $x\sim a$ and $x\sim b$. By construction of the principal value part, integrals of the first kind do not produce any divergences in the region around $x \sim a$.  Integrals of the second kind are clearly equivalent to the first kind if $a \ne b$.  So we can restrict our attention to the region $a \sim b$. 

Using the $\epsilon$ representation of the principal value part, we therefore have to evaluate
\be
\label{eq:bson1}
\CI \equiv \int dx f(x) \frac{(x-a)^2}{(x-a)^2 + \epsilon^2} \frac{(x-b)^2}{(x-b)^2 + \epsilon^2} .
\ee
We are interested in the region $a \sim b$, and also the limit of small $\epsilon$, but to see the relevant behavior we have to take a limit where $a\rightarrow b$ and $\epsilon \rightarrow 0$ simultaneously.  More precisely, we take the limit 
\be
\delta \equiv b-a \rightarrow 0, \quad y \equiv x-a \rightarrow 0, \quad \epsilon \rightarrow 0 , \qquad \frac{\delta}{\epsilon} \textrm{ and } \frac{y}{\epsilon} \textrm{ fixed} .
\ee
Then, $f(x) \rightarrow f(a)$ in the integrand, and we can integrate over $y$ from $-\infty$ to $\infty$
\be
\CI \rightarrow f(a) \int dy \frac{y (y-\delta)}{(y^2 + \epsilon^2)((y-\delta)^2+ \epsilon^2)} = f(a) \frac{2 \pi \epsilon}{\delta^2 + 4 \epsilon^2} .
\ee
The $\epsilon \rightarrow 0$ limit of the RHS is a $\delta$ function, so we have
\be
\lim_{\epsilon \rightarrow 0^+} \CI = f(a) \pi^2 \delta(b-a) + \textrm{finite} ,
\ee
as desired.

\bibliographystyle{utphys}
\bibliography{CSBib}

\end{document}